\documentclass[aps,prx,reprint,amsmath,amssymb,superscriptaddress,showpacs]{revtex4-2}
\usepackage{bm}
\usepackage{graphicx}
\usepackage[usenames, dvipsnames]{color}
\usepackage{dcolumn}
\usepackage{float}
\usepackage{hyperref}
\usepackage{eso-pic}
\usepackage{ifthen}
\usepackage{atbegshi}
\usepackage{calc}
\newcommand{\ptxt}{\color{Black}} 
\newcommand{\btxt}{\color{Black}} 
\newcommand{\angstrom}{\mbox{\normalfont\AA}}

\begin{document}

\title{Oscillatory magnetic fields for neutron resonance spin-echo spectroscopy}

\author{J. K. Jochum}
\address{Heinz Maier-Leibnitz Zentrum (MLZ), Technische Universit\"at M\"unchen, D-85748 Garching, Germany}
\email[]{Corresponding author: jjochum@frm2.tum.de}

\author{A. Hecht}
\address{Physik Department, Technische Universit\"at M\"unchen, D-85748 Garching, Germany}

\author{O. Soltwedel}
\address{Physik Department, Technische Universit\"at M\"unchen, D-85748 Garching, Germany}
\address{Institut f\"ur Festk\"orperphysik, Technische Universit\"at Darmstadt, D-64289 Darmstadt, Germany}

\author{C. Fuchs}
\address{Heinz Maier-Leibnitz Zentrum (MLZ), Technische Universit\"at M\"unchen, D-85748 Garching, Germany}

\author{J. Frank}
\address{Heinz Maier-Leibnitz Zentrum (MLZ), Technische Universit\"at M\"unchen, D-85748 Garching, Germany}

\author{E. Faulhaber}
\address{Heinz Maier-Leibnitz Zentrum (MLZ), Technische Universit\"at M\"unchen, D-85748 Garching, Germany}

\author{J. C. Leiner}
\address{Physik Department, Technische Universit\"at M\"unchen, D-85748 Garching, Germany}

\author{C. Pfleiderer}
\address{Physik Department, Technische Universit\"at M\"unchen, D-85748 Garching, Germany}

\author{C. Franz}
\address{Heinz Maier-Leibnitz Zentrum (MLZ), Technische Universit\"at M\"unchen, D-85748 Garching, Germany}
\address{Physik Department, Technische Universit\"at M\"unchen, D-85748 Garching, Germany}
\address{J\"ulich Centre for Neutron Science JCNS-MLZ, Forschungszentrum J\"ulich GmbH Outstation at MLZ FRM-II, 85747 Garching, Germany}

\date{\today}

\begin{abstract}
The generation of high frequency oscillatory magnetic fields represents a fundamental component underlying the successful implementation of neutron resonant spin-echo spectrometers, a class of instrumentation critical for the high-resolution extraction of dynamical excitations (structural and magnetic) in materials. In this paper, the setup of the resonant circuits at the longitudinal resonant spin-echo spectrometer RESEDA is described in comprehensive technical detail. We demonstrate that these circuits are capable of functioning at frequencies up to 3.6\,MHz and over a broad bandwidth down to 35\,kHz using a combination of signal generators, amplifiers, impedance matching transformers, and a carefully designed cascade of tunable capacitors and customized coils. 
\end{abstract}

\pacs{keywords: neutron spectroscopy, neutron resonance spin-echo,  MIEZE, resonant circuits}

\maketitle

\section{Introduction}\label{intro}

Conventional inelastic neutron scattering (INS) spectrometers have been highly successful in studies of both mechanical and magnetic dynamics in condensed matter systems.\cite{1972Mezei,1980Mezei} In the most basic configurations of INS spectrometers, there are intrinsic limitations arising from the inverse relationship between the amount of energy resolution and neutron flux. Neutron spin-echo (NSE) was developed primarily as a method to overcome this fundamental constraint by systematically labeling polarized neutrons with Larmor precession rates based on their individual energies (velocities). This allows neutrons from a wide wavelength band (and hence more flux from reactor and spallation neutron sources) to be utilized without losing the information fidelity required for reconstructing the energy transfers of the neutrons as they scatter from sample materials. 

The standard incarnation of NSE instruments also come with a significant limitation; the maximum energy resolution (alternatively, the range of 'spin-echo times') is directly coupled to the effective field integral $J=\int{B_0dL}$ where $B_0$ is the field strength inside a solenoid volume along the neutron path and $L$ is the path length through the field (solenoid).\cite{B_int_reference} Thus, many NSE instruments are designed with large solenoids (large radius and length along axis) to produce the maximum possible field strength over as much of the neutron path length as possible. Small variations in $J$ for different neutron trajectories causes some loss in phase information which also limits the accessible energy resolutions. Even with large radius solenoids, carefully designed correction coils (Fresnel or Pythagoras) are necessary in order to substantially minimize the variations of the field integral (for both parallel and divergent neutron paths). Successful implementation of such correction coil designs has been realized for example at the beamlines IN11 \cite{FARAGO1999270} and IN15 \cite{IN15} at ILL, which achieve field integral variations as low as $\delta J/J \approx 10^{-6}$.

Recently the J\"ulich-neutron spin-echo (J-NSE) at MLZ installed superconducting main coils, increasing its  field integral from $J$=0.48\,Tm to 1.5\,Tm.\cite{Pasini2019} In fact this reduces measurement times dramatically, because the neutron wavelength used for the same spin-echo time can be trimmed to a higher flux in the neutron spectrum. Yet the main benefit lies in the very sophisticated redesign of the coils, which decreased the intrinsic field inhomogeneties by a factor of 2.5, shifting the previous bottleneck of limiting currents in the Pythagoras-coils and thus allowing longer Fourier-times approaching 1 $\mu$s to be reached.\cite{Pasini2015} 

The technical constraints from the NSE concept outlined above may be overcome by substituting the solenoids with pairs of radio-frequency (RF) resonant spin-flippers at the boundaries of the spin-precession regions.{\btxt \cite{1992Gaehler, 2014Cook, 2016Krautloher}} {\btxt Although these resonant circuits come with their own set of technical constraints, this setup provides a way to achieve larger field integrals and smaller field inhomogeneities and thus markedly improved energy resolutions.} NSE with RF resonant flippers is conventionally referred to as Neutron Resonance spin-echo (NRSE). One of the two secondary spectrometer arms (which swivel around the sample position) is configured to make RESEDA a complete longitudinal NRSE instrument, meaning that the static $B_0$ field of the RF spin-flippers is oriented longitudinally (parallel) to the neutron beam. With the RF flippers in this longitudinal configuration, inhomogeneities in the field integral which are symmetric with respect to the solenoid axis cancel each other, and therefore substantially higher spin-echo times can be achieved.  In addition, the range of spin-echo times is limited by the maximum and minimum frequencies that can be produced by the RF flippers. However, the limitation on the low frequency end can be significantly mitigated by including elementary solenoids between the pairs of RF flippers in order to precisely tune the spin-echo time over a range exceeding seven orders of magnitude. \cite{jochum2019neutron}  

While NRSE does allow for experiments with high resolutions comparable to that of NSE with less stringent instrument design and power usage, both methods still do not allow for (in a straightforward way) the use of sample environments  which would depolarize the neutron beam. Such beam depolarization arises from magnetically ordered materials and applied magnetic fields. Furthermore, samples with strong incoherent scattering (as occurs for example in materials containing hydrogen) have a negative impact on the NRSE signal. Thus, there is a strong need for an NRSE instrument where all of the manipulations of the polarized neutron beam can be carried out before the beam reaches the sample, therefore taking away the restriction of maintaining beam polarization at the sample position. 

The so called Modulation of Intensity with Zero Effort (MIEZE) technique was developed to meet this need. \cite{1999Koppe} At the RESEDA instrument, this means that only the primary spectrometer arm is used to prepare the beam, and the secondary spectrometer arm consists only of an evacuated flight tube to prevent air scattering and a 2D time-of-flight (TOF) detector.\cite{2011Haeussler} This detector is placed at an exact position where all the various differing neutron wavelengths come together to constructively interfere, a condition which is set by a carefully chosen ratio of two different frequencies ($f_1$, $f_2$) of the RF flipper pair on the primary spectrometer arm. This process effectively takes the information encoded in polarization for conventional NRSE and effectively transfers it into a sinusoidal time-dependence of the beam intensity for MIEZE, as explained in further detail in Ref \cite{RESEDA_paper_2019}. 

\begin{figure}[tpb]
\includegraphics[width=0.95\linewidth]{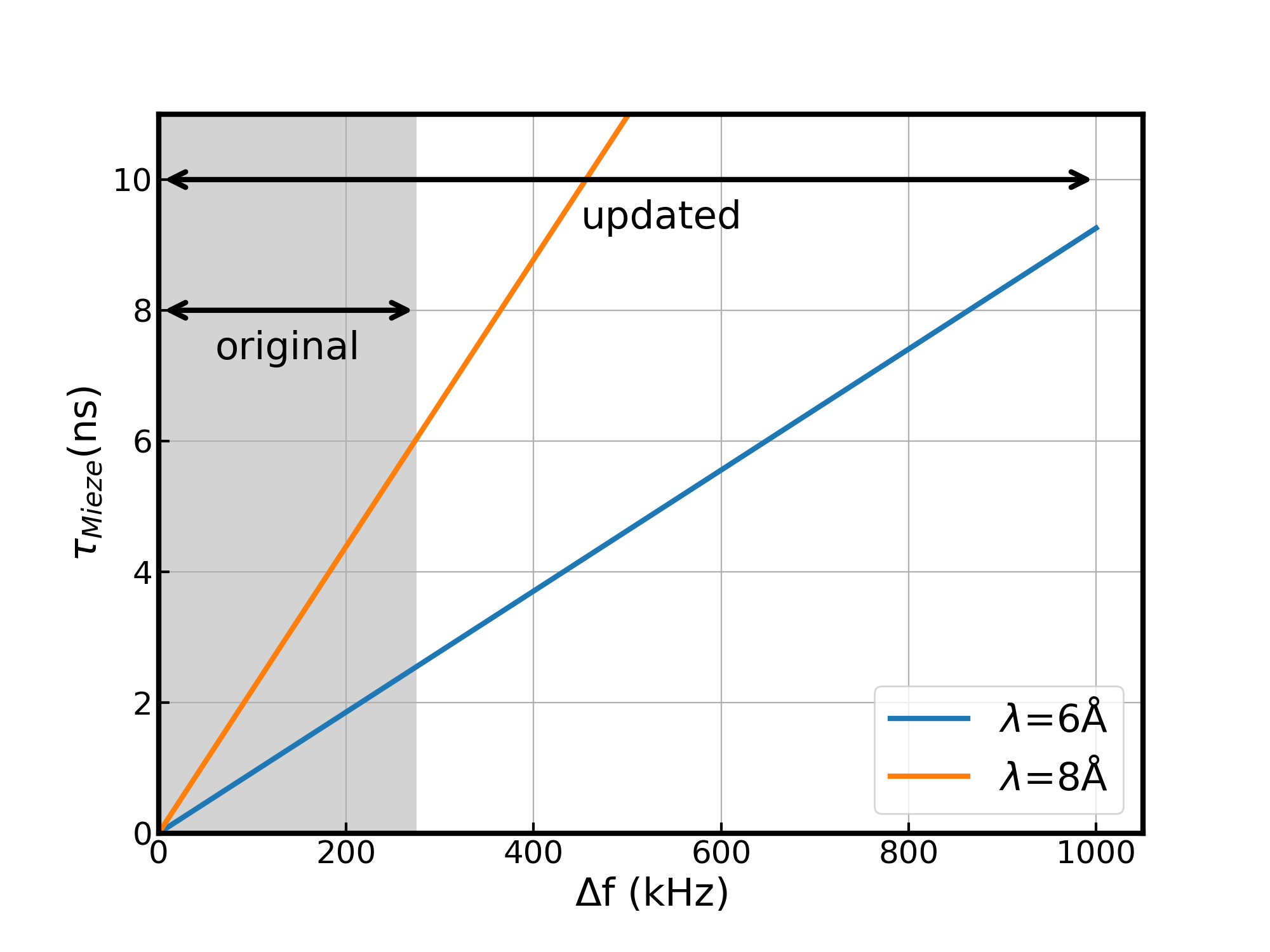}
\caption{Spin-echo time $\tau_{MIEZE}$ as a function of the frequency difference {\ptxt ($\Delta f = f_2 - f_1$, $f_{MIEZE}=2\Delta f$)} for a neutron wavelength{\ptxt s} of 6\,$\mathrm{\angstrom}$ {\ptxt and 8 \,$\mathrm{\angstrom}$} accessible at RESEDA. The resolution ($\tau_{MIEZE}$) is linear in {\ptxt $\Delta f$ but cubic in neutron wavelength $\lambda$, as can be seen in Eq. \ref{eq:taumieze}. The shaded area corresponds to the ``originally'' available range of $\Delta f$ before the ``updated'' resonant circuits were installed, where $\Delta f$ now spans from 0.1\,kHz to 1000\,kHz.}} 
\label{tvf_fig}
\end{figure}

{\ptxt Within the framework of the spin-echo approximation (small energy transfers), the difference $\Delta f = f_2 - f_1$ relates to the spin-echo time, $\tau_{MIEZE}$, via:

\begin{equation}
	\tau_{MIEZE} = \frac{2 m_n^2}{h^2} \lambda^3 L_{sd} \Delta f 
	\label{eq:taumieze}
\end{equation}
where $m_n$ is the neutron mass, $h$ is Planck’s constant, $L_{sd}$ is the sample-detector distance, and $\lambda$ is the neutron wavelength. This relationship is plotted in Fig. \ref{tvf_fig}, showing the linear dependence of $\tau_{MIEZE}$ (which is proportional to energy resolution) on $\Delta f$. The achievable $\Delta f$ depends on the range of frequencies accessible to the resonating circuits.  The lower limit is determined by the Bloch-Siegert shift (35\,kHz) \cite{1940Bloch}, while the upper limit (3.6\,MHz) is set by technical constraints as described in further detail in the following discussion. }

The ability with NRSE and MIEZE to measure the equivalent of the intermediate scattering function $S(Q,\tau)$ (where $Q$ is the neutron momentum transfer and $\tau$ is the spin-echo (or $\tau_{MIEZE}$) time) over an exceptionally wide dynamic range allows for a commensurately wide scope of dynamical processes to be measured in a single experiment. As an example, MIEZE measurements with the pyrochlore system Ho$_2$Ti$_2$O$_7$ (known to host a spin-ice state with diffusing magnetic monopoles at low temperatures \cite{Castelnovo2008}), were able to resolve the slow spin dynamics in order to uncover the magneto-elastic coupling of phononic modes to crystal field transitions. Here the phonon coupling to the crystal field is essential to model the measured $S(Q,\tau)$ over the full range of MIEZE time $\tau$. \cite{wendl2018neutron}. MIEZE has also allowed studies of the dynamics (i.e fluctuation timescales) of magnetic skyrmions condensing from a topologically trivial paramagnetic state in an applied magnetic field. This study was able to prove that the lifetimes of these skyrmion lattice fluctuations is on the order of nanoseconds, which may be considered quite long in this context. \cite{Kindervater_PRX_2019}

By pushing beyond the previous high frequency limit of $f$\,=\,1\,MHz for the resonating circuits in NRSE and MIEZE, further observations of a wider variety of dynamical processes has been made possible at RESEDA. In this paper, we provide the precise technical requirements to produce an NRSE and/or a MIEZE instrument with these capabilities. With these RF-circuits now capable of reaching $f$\,=\,3.6\,MHz and the future availability of the appropriate static $B_0$ strength, the NRSE arm at RESEDA is expected to achieve up to 50\% of the field integral of the world leading NSE spectrometers. Furthermore, in MIEZE mode with experiments set-up in a SANS configuration, measurements with a $\tau_{MIEZE}$ of tens of nanoseconds can now be achieved. 

\section{Theoretical Considerations}

At the heart of neutron resonance spin-echo spectroscopy are resonant neutron spin flippers. They are comprised of a static magnetic field imposing the Larmor precessions of the neutron spin and a rotating field matching that precession velocity \cite{Martin2014}. Technically the rotating field is replaced by an oscillating field, which can be seen as a superposition of two counter-rotating fields. The opposing direction may be neglected for high enough frequencies \cite{1940Bloch}, in the case of RESEDA $\geq$\,35\,kHz. There a several possibilites to create such a rotating field, and the choice depends on spectrometer type and desired conditions to meet.

In the case of RESEDA, a longitudinal NRSE spectrometer at a continuous reactor neutron source, a good homogeneity over the beam cross section and a sharp field boundary in neutron flight direction are essential. The available space is restricted by the gap between the Helmholtz $B_0$ coils, typically $\sim$\,2\,cm. Therefore, the RF flipper coil must be designed to have specific electrical properties to account for this; namely a resistance of 0.5-1\,$\Omega$ and inductance of 20-30\,$\mu H$. To reach currents of 5\,A at a frequency band from 35\,kHz to 4\,MHz, a resonant circuit may be used. For other geometries of the instrument, e.g. the versatile beamline LARMOR at the ISIS spallation neutron source, a simpler design of the RF coils may be operated with a combination of a frequency generator and a powerful amplifier \cite{Larmor2019}. Additionally, in recent years superconducting materials have been utilized in Wollaston prisms \cite{Wollaston2014, Wollaston2017} and resonant circuits \cite{SC_RF_flipper} to achieve resonant spin flips in Larmor labeling techniques.

Resonating circuits have been used for a long time in conventional NRSE instruments, however, the maximum frequency was always limited to below 1\,MHz \cite{2014Cook}. In the following we will briefly review the reasons for this and potential limitations. 

For transverse NRSE, the RF coil is enclosed in the B$_0$ coil providing the static magnetic field. This close vicinity with a gap of less 1\,cm leads to both a shift in the resonance frequency to lower values and a damping of the amplitude. In LNRSE however, this problem is solved naturally by the design of the Helmholtz coils. Particularly important is the choice of materials, both for the actual coil and the components of the resonating circuit. For the coil, the skin of a single wire and the proximity effect between neighboring wires must be considered. Both are reducing the usable cross section of the wires for the conduction of AC currents with increasing frequency, and therefore increase the Ohmic resistance of the resonance circuit. Furthermore, all electronic components, such as relays, add parasitic capacities to the resonant circuit and therefore reduce the maximum achievable frequency. Special attention has to be paid to the cables connecting the individual components of the resonating circuit. Different types of coaxial cables and BNC connectors contribute to the impedance of the circuit and add damping. 

Another important issue is the inter-winding capacitance, which together with the coil inductance acts as a parallel resonant circuit. The resonance frequency of this unwanted circuit has to be well-separated from the resonance of the serial circuit, otherwise it will significantly increase the resistance and dampen the amplitude of the resonance.


\section{Technical Realization}


In the following section the technical realization of the resonating circuits at RESEDA is described in detail. 
{\ptxt The schematics of the circuit boards for C-boxes 1 and 2 can be found in the Appendix (Section \ref{app}), while the KiCAD files, including the .gbr files needed to print the boards are published in \cite{figshare}.}

\begin{figure}[tbp]
\includegraphics[width=\linewidth]{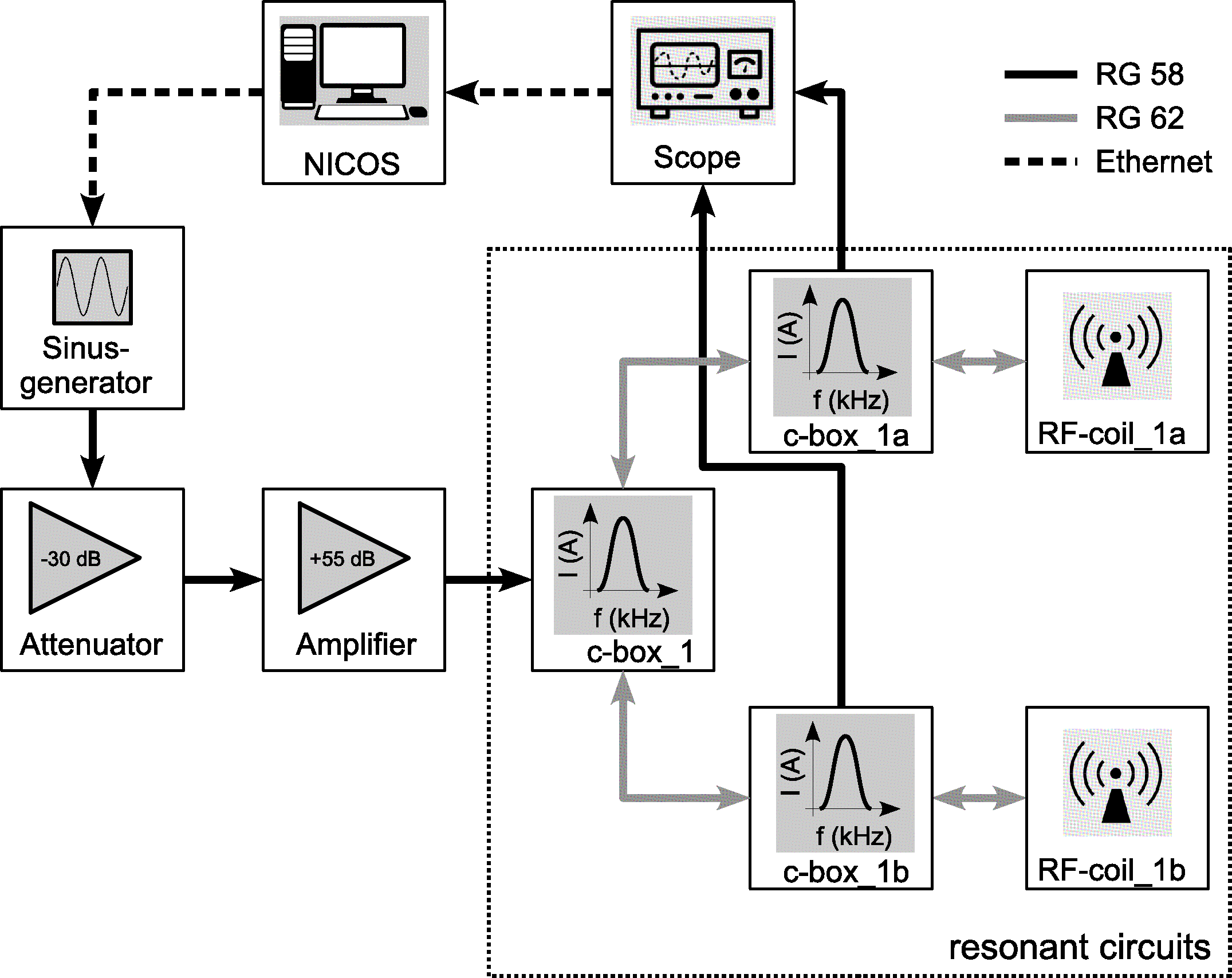}
\caption{Schematic circuit diagram of the resonating circuit with two coils per circuit (NRSE configuration).}
\label{pic:plan}
\end{figure}

The resonating circuit is depicted schematically in Fig. \ref{pic:plan}. The circuit consists of a signal generator which feeds into an amplifier, via a -30\,dB attenuator.
The signal then connects to a matching circuit. The matching circuit is split into two parts, one part for frequencies below 1\,MHz (C-box 1), and a second part for the higher frequencies (C-box 2). The latter is fixed directly at the coils to reduce the length of the cables between the C-box and the RF coil to an absolute minimum. The matching circuit is connected to the RF coils as well as to an oscilloscope providing feedback to the operating software NICOS \cite{nicos} to allow for fast regulation.

In the present implementation of RESEDA, at the primary spectrometer arm two separate circuits are connected to a single RF coil each. A third circuit controlling two RF coils is installed on the secondary spectrometer arm, used for NRSE. In this configuration it is possible to switch between NRSE and MIEZE within a few minutes while keeping changes of the necessary technical equipment and reconfigurations to a minimum. The different parts of these circuits are described in the following sections.

\subsection{Signal generation and amplification}

The sinusoidal signals for the RF circuits are supplied by Agilent HP 33220A signal generators, that offer the desired frequency band. Depending on output frequency and level their harmonic distortion is below -35\,dBc. 

Since the amplitude regulation of the RF circuits is driven by the amplitude of the generated signals it is advantageous to match them to the amplifiers (Rohde \& Schwarz BBA150-A125) nominal input power (-3.4\,dBm into 50\,$\Omega$ which equals 0.427\,mV). 
Using -30\,dB (factor of 31.6 for voltage) attenuators allows the use of the full output range of the signal generators ($V_{pp}$ = 10\,mV\,...10\,V into 50\,$\Omega$) and makes it possible to use them for precise amplitude regulation. Additionally it prevents the input stage of the BBA150s from any harmful signal intensity levels and more importantly, shifts the minimal achievable power output from 3\,W down to 0.1\,W. This is crucial to tune the current in the low Ohmic RF coils (0.5\,$\Omega$, cf. figure \ref{fig:omega}) down to 500\,mA (needed current for an RF-$\pi$-flip at $\lambda\,=\,15\,$\AA) at low frequencies.

\subsection{Matching circuit for the new RF coil}

To suppress distortions of the RF signal and to maximize the efficiency of the circuit \cite{2014Cook}, the resonant circuit needs to be purely resistive, and any impedances will need to be matched by a matching circuit. At RESEDA this was done with a toroidal transformer and a serial resonance. 
The 50\,$\Omega$ output impedance of the amplifier was transformed to the corresponding serial resistance of the RF coil using a toroidal transformer. A series capacitor compensates for the reactance of the coil. 

The matching circuit contains frequency dependent series resistances, for which different transformers were required. Since it was possible to implement two transformers on one toroid this represented a minor issue. 

RG62 cables (93\,$\Omega$) were used to connect the coil to the matching circuit, due to their small capacity of 44\,pF/m . Standard 50\,$\Omega$ coaxial cables would have limited the use of the circuit to frequencies below 1.5\,MHz. 

\subsubsection{Matching capacities}
{\ptxt The matching capacities were divided into a low frequency part (C-box\,1) and a high frequency part (C-box\,2). This division makes it possible to keep the cables (RG62) connecting C-box\,2 to the RF - coil as short as possible, minimizing its impedance. For this C-box\,2, which contains all high-frequency components, is placed directly on the aluminum support of the coil (Figure \ref{fig:RFC} item (v)). }

The main components of C-box\,1 are the capacitors and the transformer, as described in the following. Additionally, C-box 2 includes a pick-up coil/current transformer as well as high frequency relays and capacitors. To illustrate, the internal components of C-box\,2 are shown in Fig. \ref{fig:cbox2}. 

C-box\,1 is controlled using the software NICOS \cite{nicos} via an Ethernet connection. It receives its input signal directly from the R\&S amplifiers. The output of C-box\,1 is connected directly to the RF coil for frequencies below 1\,MHz. For frequencies above 1\,MHz, the output signal is routed via C-box 2.
\\

The capacitances required for tuning {\ptxt the resonating circuits} need to cover a large range from 1\,$\mu$F\,-\,50\,pF under voltages up to 3\,kV. Furthermore, to reduce power losses a high quality factor is needed under currents up to 5\,A. To fulfill these requirements three sets of capacitors are used covering the entire frequency range of the resonating circuit:

\begin{itemize}
    \item f $<$ 300\,kHz: film capacitors, WIMA MKP10 and FKP1 {\ptxt (C-box\,1)}
    \item 300\,kHz $\leq$ f $\leq$ 600\,kHz: mica capacitors, Cornell Dubilier, CDV16 {\ptxt (C-box\,1)}
    \item f $>$ 600\,kHz: high voltage ceramic capacitor, AVX, HQCE and HQCC (C-box\,2)
\end{itemize}
The relays need to have low transmission losses and withstand currents larger than 5\,A and, depending on frequency effective voltages up to 2\,kV. For frequencies larger than 1\,MHz the parasitic capacitances of the relays need to be as low as possible when the relay is switched off.  Therefore, similarly to the capacitors, individual relays are used for different frequency ranges:
\begin{itemize}
    \item f $\leq$ 1\,MHz: PCB relays, Finder, 40/41 {\ptxt C-box\,1}
    \item f $>$ 1\,MHz: high voltage reed relays, GIGAVAC, G41A ({\ptxt C-box\,2, }see Fig. \ref{fig:cbox2} (iv))
\end{itemize}

\subsubsection{Impedance matching}

It is essential that the transformer lowers the output impedance of the amplifier from 50\,$\Omega$ to 0.6\,-\,6\,$\Omega$. 
Due to this large transmission ratio, 
thin copper tape (0.05\,mm x 5-10\,mm) is used as a secondary winding. The winding width of the two windings are then almost identical, maximizing the coupling between the two. 

The transformers use a FT250-75 ferrite core. 
Three transformer ratios are required to cover the full frequency bandwidth: 50:1\,$\Omega$, 50:2.2\,$\Omega$, 50:4.6\,$\Omega$. For all three, the secondary winding consists of three windings of flat copper wire, with widths of 10, 7 and 5\,mm respectively. The primary winding consists of 0.7\,mm thick enamelled copper wire, with winding numbers of 21, 15 and 10. The higher the ratio, the higher the winding number.  
Due to their induction leakage, the transformers also effectively act as a low pass filter and cause a phase shift of maximum of 20\,$^o$ at the resonance frequency only. Thus we omit a low pass compensation.\\

\subsubsection{Signal shaping electronics}

The matching circuits were designed to work with real (i.e. non-ideal) signals, which contain for example higher harmonics from class AB amplifiers that can be reflected by the resonant circuits. Therefore the incoming signals are preemptively filtered. The switchable diplexer passes signals with frequencies below 5\,MHz via the low pass, while frequencies above 5\,MHz are adsorbed in the resistor in the high-pass filter. Additionally a variable high pass filter terminates high frequency signals above 0.27 to 21\,MHz subject to specific requirements. That is, boxes allow for the operation of two individual rf-coils using one amplifier (and signal generator). Therefore a power divider feeds two identical matching boards.

\subsection{Pick-up coil}

\begin{figure}[tpb]
\includegraphics[width=\linewidth]{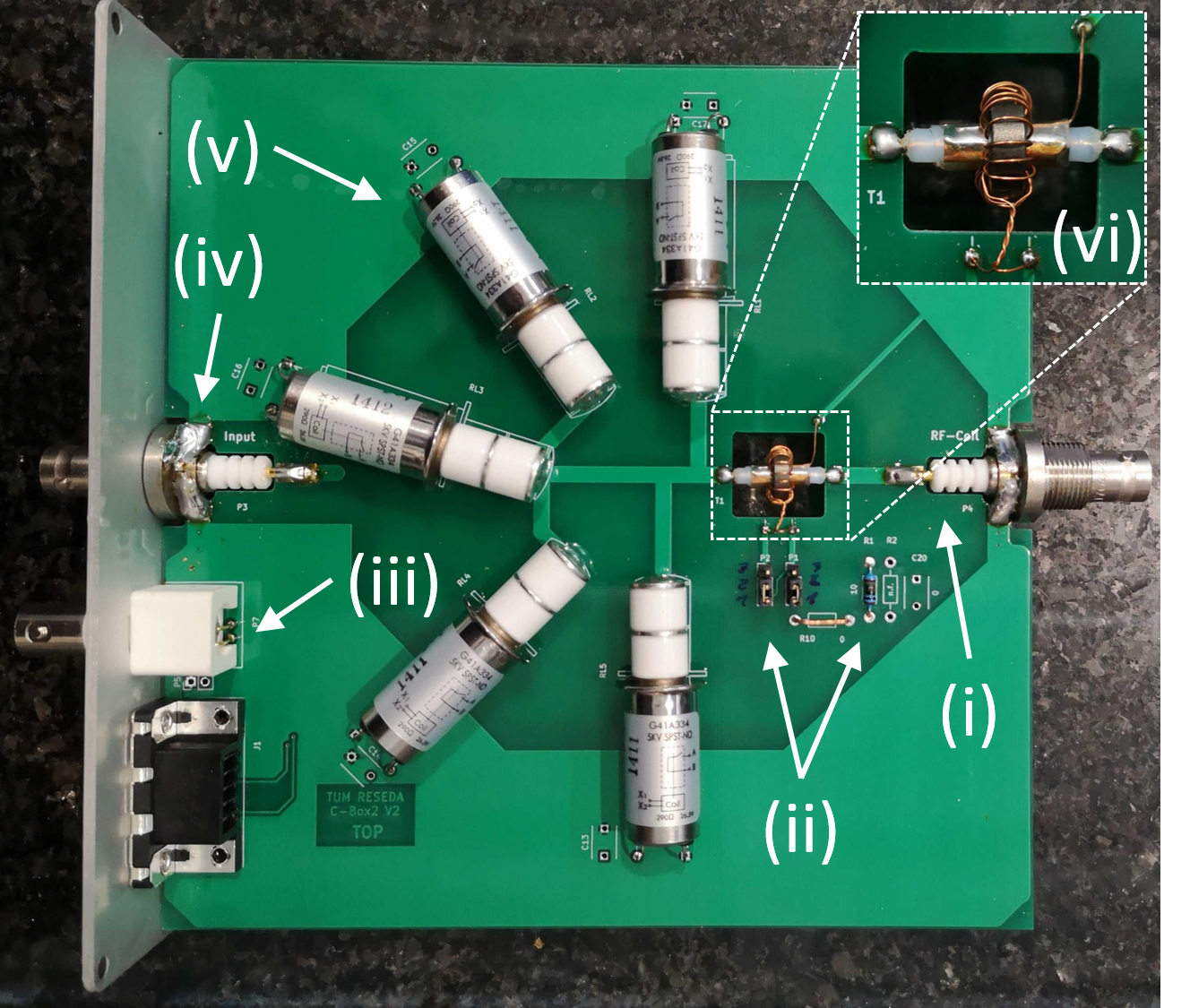}
\caption{C-box 2 currently in use at RESEDA. (i) Connection to RF coil (see Fig. \ref{fig:RFC}) (ii) Resistances (iii) Connection to the oscilloscope  (iv) Connection to C-box 1 (v) High voltage reed relay (vi) Pick-up coil.  }
\label{fig:cbox2}
\end{figure}

\begin{figure}[tbp]
\includegraphics[width=0.95\linewidth]{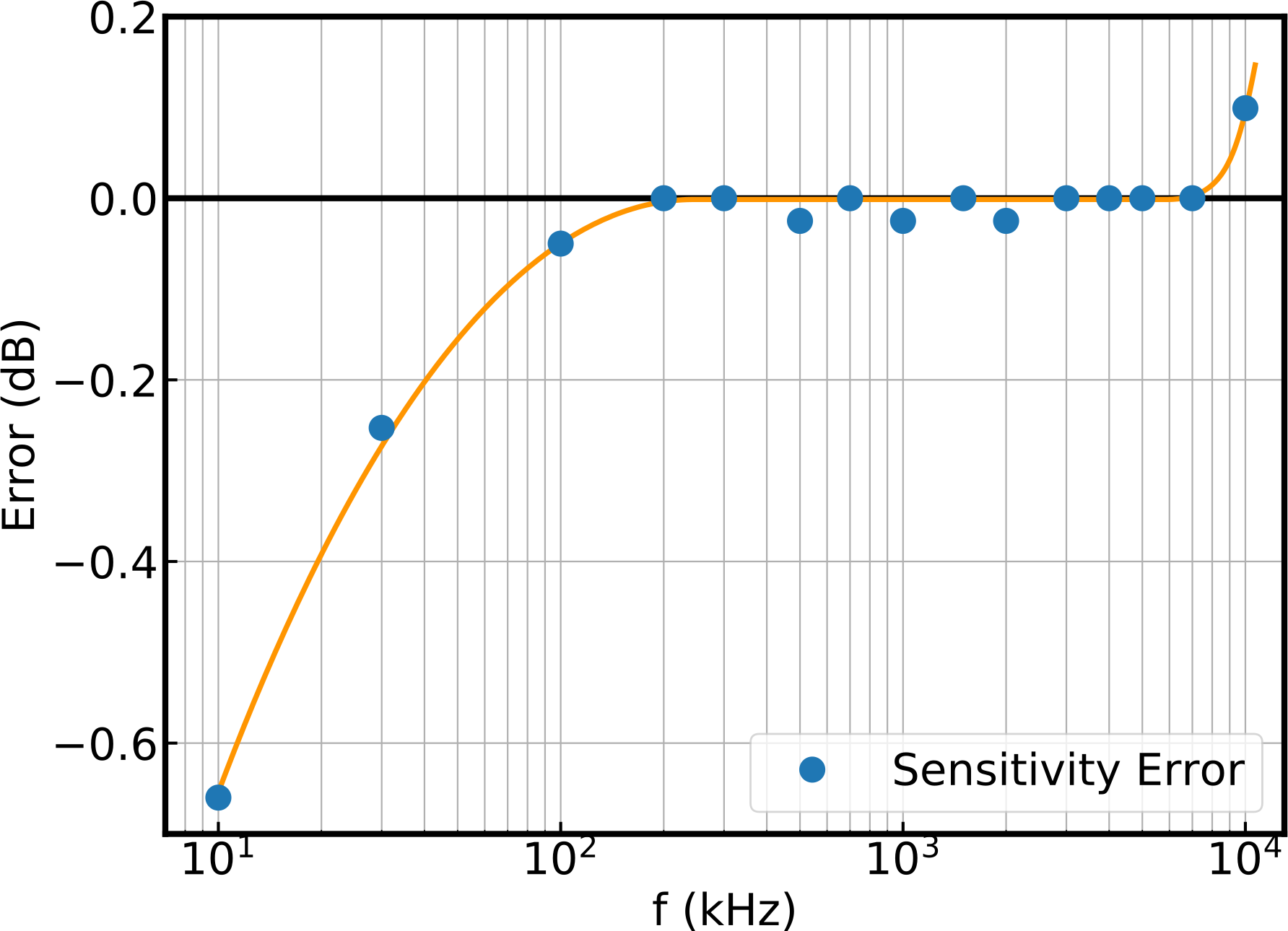}
\caption{Error of the readout coil in dB versus frequency. The orange line serves as a guide to the eye. The error is close to zero in the desired range of 25\,kHz to 4\,MHz, with only slight deviations of -0.25dB at 30\,kHz.}
\label{fig:Sens}
\end{figure}

The sensing elements for the feedback loops (the pick-up coils) have to operate at frequencies up to 4\,MHz. To obtain constant transformer characteristics over the entire frequency band a FT37-72 core ($\mu_r$ = 2200) with a winding number of 20 is used. 
The terminating load of the secondary resistor is 10\,$\Omega$, making the real part of the forward resistance smaller than 50\,m$\Omega$.
Because the voltage at the pick-up coil is {\btxt roughly 1\,kV} for frequencies above 1\,MHz, cross talk between the capacitors and the secondary winding {\btxt is an issue. As an estimate: With a capacity of 1\,pF the resulting current in the secondary circuit is 25\,mA at 4\,MHz which is on the same order as the secondary current generated by induction (50\,mA).
To suppress capacitive cross-talking between the primary and secondary circuit, a thin grounded copper foil is introduced as an insulator (shown in the inset of Fig. \ref{fig:cbox2}). Indeed the sensitivity over the full frequency range is nearly constant with negligible deviation as shown in Fig. \ref{fig:Sens}. }\\

\subsection{The RF coil}

\begin{figure}[tbp]
\includegraphics[width=\linewidth]{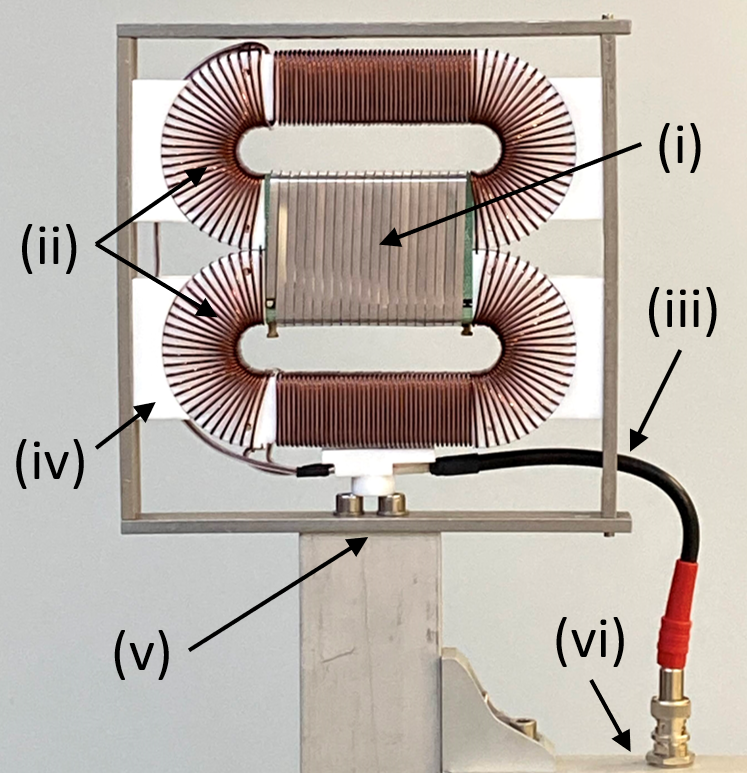}
\caption{Radio frequency coil currently in use at RESEDA. (i) Main coil (ii) Compensation coils (iii) High voltage connector (iv) Teflon holders (v) Aluminum support and frame \mbox{(vi) C-box 2}.}
\label{fig:RFC}
\end{figure}

The RF coil represents the key element of the resonating circuit, responsible for the optimal operation of the entire instrument. The RF coil consists of a main coil and two compensation coils (see Figure \ref{fig:RFC}). The compensation coils are connected in parallel to each other and in series to the main coil {\ptxt \cite{Martin2014}}. Their purpose is to compensate for the stray fields produced outside the main coils in the path of the polarized neutron beam, thereby reducing power losses from induced currents in the adjacent metallic structures (such as the static field coil) and preventing perturbations of the static field \cite{2014Cook}.  

Previously, the radio frequency coils (RF coils) at RESEDA were operating up to a resonant frequency of 1\,MHz. 
As mentioned in Section \ref{intro}, the possible spin-echo times ($\tau_{MIEZE}$) that may be reached with a MIEZE spectrometer strongly depend on the frequencies that can be achieved with the RF flippers. To access higher $\tau_{MIEZE}$, it is paramount to optimize the resonating circuits. The static field coils available at RESEDA permit resonant frequencies up to 4\,MHz. There are several difficulties in achieving such frequencies: 
\begin{itemize}
    \item The relatively high inductance of the RF coils (between 20 and 30\,$\mu$H) requires very small capacities (several pF) which can only reliably be achieved by avoiding all parasitic capacities.
    \item At frequencies above 1\,MHz the skin effect leads to a strong increase in Ohmic resistance of the RF coils which in turn dampens the resonant circuit, making it increasingly difficult to feed power into the coil. Furthermore, the increase in resistance leads to a warming of the coil.
\end{itemize}

\subsubsection{The main coil}
The main coil is wound from aluminum flat wire (0.2\,mm x 3\,mm, 19 windings) and has an inductance of 6.4\,$\mu$H, contributing less than 1/3 of the inductance of the entire coil (L$_{coil}$ = 23.8\,$\mu$H).\\

At a neutron wavelength of 4.5\,\AA\ the maximum current running through the coil will be 4.4\,A.
{\ptxt This leads to a power loss of  9.68 $R_S$, where R$_S$ is the serial resistance.} Keeping the power loss at a maximum of 50\,W this limits the Ohmic losses of the coil at a frequency of 4\,MHz to a maximum allowed resistance of 5\,$\Omega$. 

\subsubsection{The compensation coils}

As may be seen from Fig. \ref{fig:RFC} the compensation coil consists of two toroidal half-sections with a straight section in between.

The compensation coils are the main contributors to the Ohmic losses of the RF coil. Therefore it was necessary to optimize their design to match a frequency of 4\,MHz. 

The resistance of the compensation coil can be minimized by using a solid copper wire (diameter = 0.63\,mm).
Below frequencies of 1\,MHz, the resistance of the multistrand (Litz) wire is lower than that of the single strand wire due to a reduction in skin- and proximity effects. Above frequencies of 1\,MHz, the increase in resistance is much stronger for the multistrand wire than for the single strand wire due to the parasitic capacitance between the different strands.

The toroidal part of the plastic core for the compensation coils was constructed with grooves to accommodate the wires so that they remain in place. 
However, the grooves are only placed at the outside part of the toroidal core, since otherwise the wires would be packed very tightly on the inside, increasing the resistance.

\begin{figure}[tbp]
\includegraphics[width=1.01\linewidth]{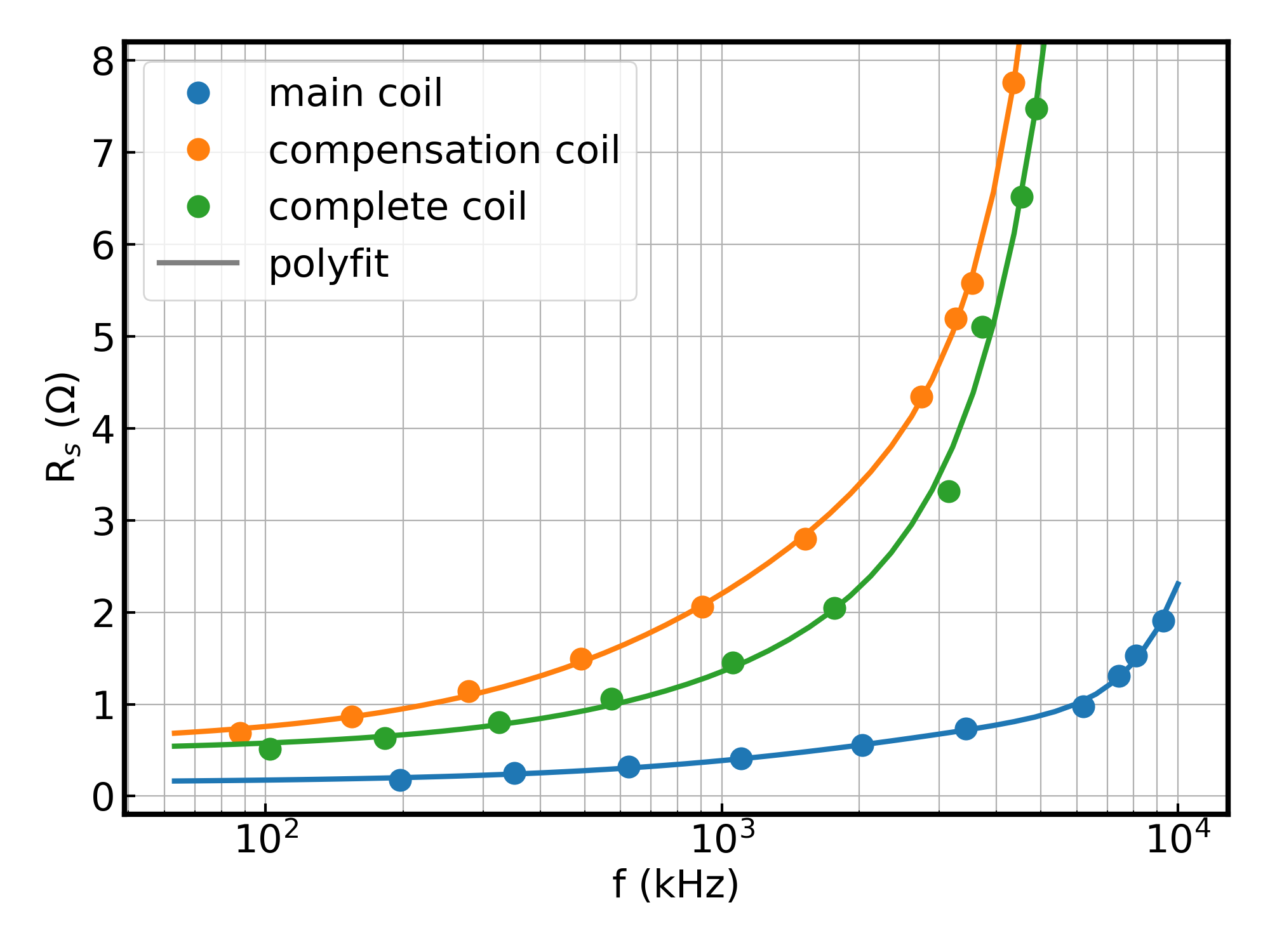}
\caption{Resistance of the main coil, compensation coil, and the complete RF coil currently installed at RESEDA as a function of frequency $f$. The resistance drastically increases with higher frequencies. The solid lines represent cubic fits.}
\label{fig:omega}
\end{figure}

With these specifications the resistance of the compensation coils is 7.0\,$\Omega$ for $f$ = 4\,MHz, while the resistance of the entire RF coil is 5.0\,$\Omega$ at the same frequency.
Figure \ref{fig:omega} shows the resistance of the main coil, the compensation coil, and the RF coil as a function of frequency $f$. The figure clearly shows the extent to which the compensation coils are the primary contributors to the Ohmic losses. \\

Taking into account the DC resistance of the RF coil of 0.49\,$\Omega$, its inductance of 23.8\,$\mu$H, and parasitic capacity (connectors and winding's) then the parasitic resonance is at 10.4\,MHz (in parallel with its primary) for a distance of 73\,mm to ground. This is more than a factor of two above the current maximum operating frequency of 3.6\,MHz, and thus suppresses interference effects for all operational settings of the instrument.

\subsection{Control loop of the RF circuit in NICOS}

For a reliable resonant spin-echo setup the magnetic RF-fields have to exhibit long-term stability at values that may be set accurately in a highly reproducible manner. Since parameters leading to instabilities may be of different origin, we implemented closed RF-current control loops for each RF circuit. This allowed us to ensure the optimal magnetic field strength for RF $\pi$-flips with superior efficiency, after taking into account the individual electric characteristics, varying environmental conditions, and different possible choices of circuit frequency or neutron wavelength (velocity). The layout is depicted schematically in Fig. \ref{pic:plan}. The actual controller is embedded digitally in the NICOS control software \cite{nicos}. 

Since the current is directly proportional to the magnetic field strength and an easy variable to change, it was chosen as the loop process variable. The current is generated via induction in the pick-up coils inside C-box 2. The sensitivity of the pick-up coils for the different resonating circuits is practically identical and more importantly, effectively constant over the desired frequency band (as specified in Section II A) allowing the feedback parameters for all working conditions to be fixed. 

A four channel oscilloscope (50\,$\Omega$ input impedance, 1\,GHz bandwidth, 16\,bit vertical resolution, RTE1104, Rohde \& Schwarz, Germany) serves as an analog to digital converter. The internal FFT of the oscilloscope allowed real time evaluation of all signals,  including the values of power in individual channels for frequencies spanning two orders of magnitude (35\,kHz to  4\,MHz). To accomplish this, the time base of the oscilloscope was adjusted according to $\frac{100}{f_2}\frac{\text{sec}}{\text{div}}$  whenever a new spin-echo time $\tau$ was set. 

Of course, one might optimize the amplitude digitization (voltage resolution) as well, but this value differed only by a factor of three at most ($\approx$550\,mV at 15\,\AA~vs. 1.5\,V at 4.5\,\AA). Having a 16-bit analog-to-digital converter (ADC) at a voltage range of 4000\,mV, i.e. 400\,$\frac{\text{mV}}{\text{div}}$ and thus covering all expected voltages, resulted in a voltage resolution of 0.061\,mV for the ADC. This resolution is more than 10 times better than the accuracy when tuning the excitation amplitude of the sine (signal) generators, thus allowing the vertical deflection of the oscilloscope to be kept constant. 

In order to tune the RF circuit amplitude, the excitation level of the sine generator was varied by utilizing a simple proportional control loop with a dead-band, i.e. no action occurring below a set input, of $\pm$50\,mV around the set-value and limited stepping values in a range between 1 and 100\,mV. This tuning procedure yields reproducible and stable currents.

The polling interval of the read-out was $\sim$ 1\,s. This limited the dynamical loop characteristic, namely the time required to change the RF-amplitude, to a transient time of approximately 10\,s. Compared to the typical counting times ranging from 300 to 3000\,s per spin-echo point, the dynamical loop characteristic is still two orders of magnitude faster. This means a dead-time below 3\% of the counting time when setting new spin-echo times.


\section{Characterization of the new resonating circuits}

\begin{figure}[tbp]
\includegraphics[width=0.98\linewidth]{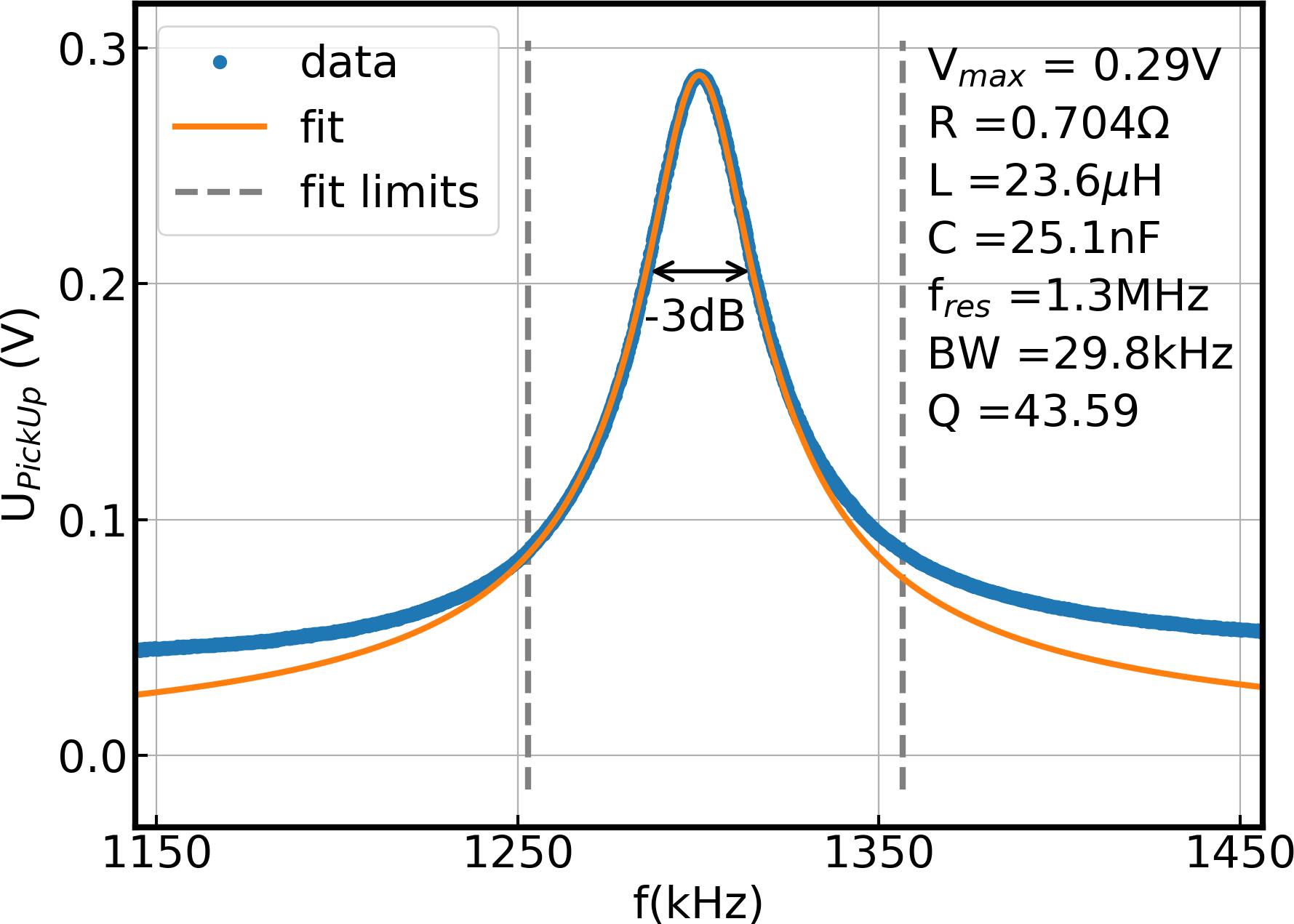}
\caption{Resonance scan of a rf-flipper coil around resonance frequency f\,=\,1.301\,MHz with frequency generator set to 1\,V output. Diplexer was set on, highpass off,    Parameters are derived by fitting (orange line) the data (blue symbols) using Eqn. \ref{R_curve_equation}.}
\label{pic:sfscan}
\end{figure}

\begin{figure}[tbp]
\includegraphics[width=\linewidth]{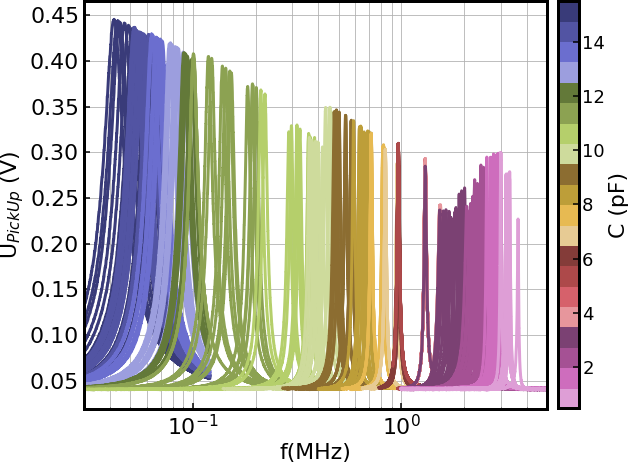}
\caption{Frequency scans for various setting of the capacitor box at 1\,V amplitude at the frequency generator and fixed gain at the amplifier. Each color represents one scan at a setting.}
\label{pic:fscan}
\end{figure}

\subsection{Resonances and resonance band}

In this section, a thorough characterization of these resonant circuits is presented. The experiments were performed with a free standing coil at the RESEDA instrument. The setup characterized here is now in operation for instrument users. As described above, operation of the devices and data recording is done using the NICOS instrument control software. 

To determine the resonance frequency for a given set of capacities, the frequency of the signal generator was scanned while holding the output voltage constant at a low value of 1\,V with a fixed amplifier gain. The rms voltage at the pick-up coil, shown in Fig. \ref{pic:sfscan}, was recorded using the oscilloscope. The peak maximum indicates the resonance frequency, in this example 1.3\,MHz. The resonance curve may be fitted with a simple Lorentzian,
\begin{equation}
I \sim V_{pick-up} = \frac{V_{max}R}{\sqrt{R^2 + (f \cdot L - 1/(f \cdot C))^2}}
\label{R_curve_equation}
\end{equation}
where $R$ is the resistance, $L$ the coil inductance, $C$ the capacitance, $f_{res}$ the resonance frequency, BW the bandwidth at -3\,dB indicated by the black arrow, and $Q$ is the quality factor. The fit is carried out within the dashed grey lines, to avoid the base noise level of $\sim$0.05\,V.

To find the available frequency band, a frequency scan was performed for various possible combinations of capacitors with appropriate transformer and filter settings. 
Figure \ref{pic:fscan} shows the resulting resonances ranging from 35\,kHz to 3.6\,MHz. 
Lower frequencies are technically possible, though not useful at RESEDA due to the Bloch-Siegert shift \cite{1940Bloch}, which decreases the flipping efficiency at low frequencies. 
To achieve a high dynamic range of resolutions, the field subtraction method is used instead \cite{jochum2019neutron}. 
In the high frequency regime, the current setup was tested for resonances up to $f$\,=\,3.6\,MHz, just below the $f$\,=\,4\,MHz value that allows for the highest instrument resolution with the current static field coils. 
We find several smaller gaps in the frequency band on account of the logarithmic spacing of the values of commercially available capacitors used in this setup. 
This limitation may in principle be overcome by using custom-made capacitors with values stepped in multiples of two. 
For operation at the instrument, only the gap between 1-1.3\,MHz proves to be inconvenient. The amplitudes decrease slightly with increasing frequency as the resistance of the circuit increases with frequency. 
When the transformer ratio is changed, the amplitude steps are found to increase.
However, the amplitude is high enough to allow for a $\pi$-flip of neutrons even at the lowest neutron wavelength of 4.5\,\AA\, where the highest current is needed.

\subsection{High-frequency resonant neutron spin flip} 

To perform a resonant spin flip at the highest available resonance of 3.6\,MHz, a high static field of the $B_0$ coils is needed. 
In addition, due to the proximity of the RF and static field coils, the static field required to perform a spin flip at 3.6\,MHz is increased beyond the limits ($\sim$150 mT maximum field) of the presently installed static field coils. {\btxt Therefore, we temporarily installed 
a MLZ sample environment magnet (equipped with a holder for RESEDA's RF coil) with a sufficiently large 80 mm gap between its superconducting Helmholtz coils which leaves enough space between the RF and static field coils. With this gap, a B$_0$ field of only 122\,mT is required as evidenced in the following description.} 

\begin{figure}[tbp]
\includegraphics[width=0.98\linewidth]{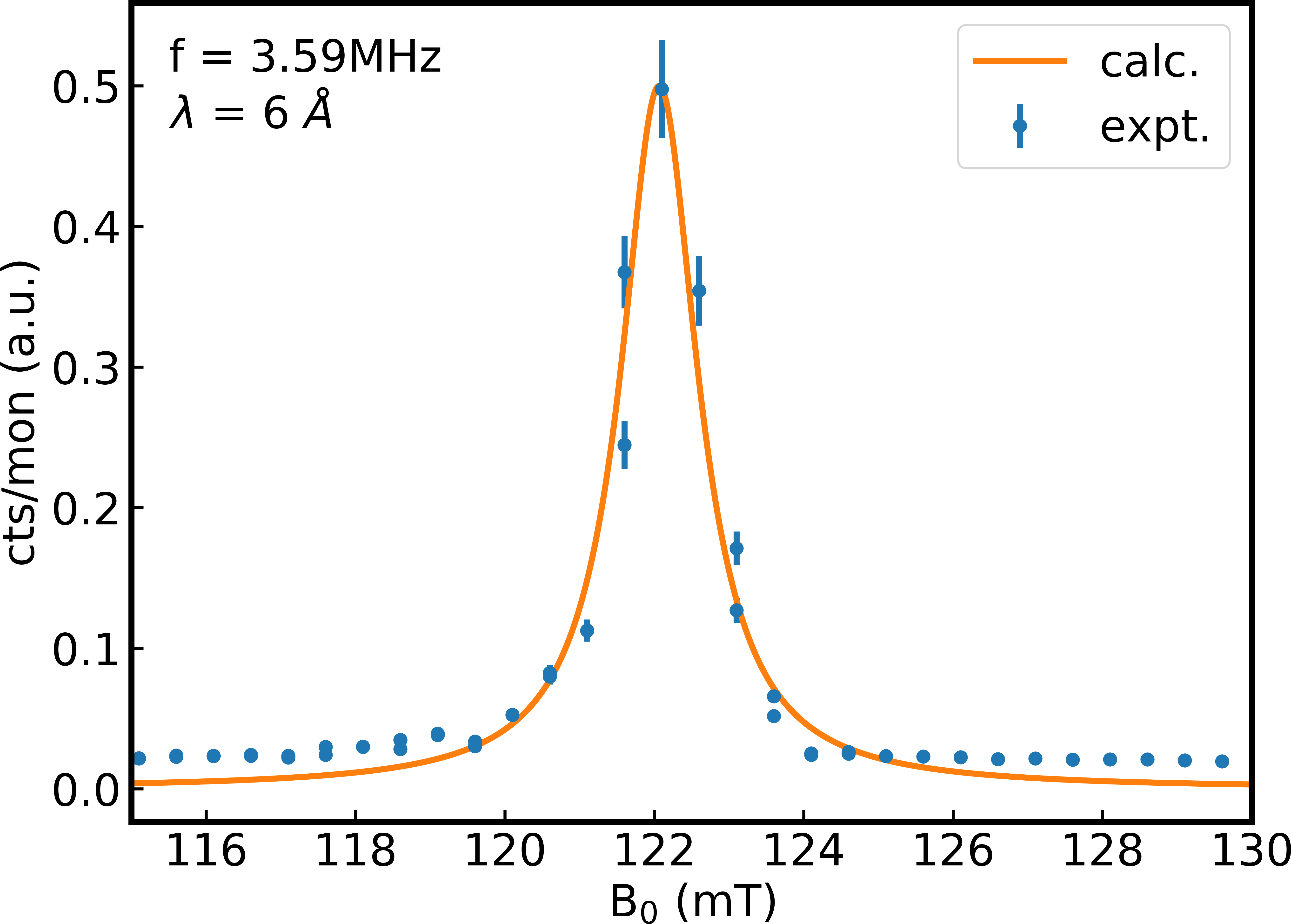}
\caption{Scan of static magnetic B$_0$ field at a frequency of $\omega_{rf}$\,=\,3.6\,MHz. The peak indicates a resonant flip with a efficiency of $>$\,92\,\%.}
\label{pic:bscan}
\end{figure}

In this test experiment the polarization of the neutron beam was maintained by the instrument's guide field, and flipped by the combination of the sample ($B_0$) field and the RF flipper at the sample position. The polarization was then analyzed by an S-bender \cite{Sbender_2006} and the neutron counts were recorded by a single $^3$He counter tube. The amplitude of the resonating field is fixed at the value needed for 6\,\angstrom\ neutrons while the $B_0$ field was scanned (c.f. Fig. \ref{pic:bscan})

The clear peak in the data {\btxt in Fig. \ref{pic:bscan}} above 122\,mT corresponds to a $\pi$-flip, {\btxt demonstrating that it is possible to achieve a resonant spin flip at 3.6\,MHz for neutrons with a wavelength of 6\,\angstrom.} The data quality was limited as the scan was performed at the end of the reactor cycle when the neutron flux was decreasing. Good agreement with the calculated curve is observed. This proves the possibility to perform RF-flips at very high frequencies as a precondition to enhance the resolution of NRSE and MIEZE drastically.


\section{Conclusion}

{\btxt \subsection{Summary}}

We have designed, constructed, and tested new resonating circuits including next generation RF coils for the NRSE and MIEZE options of the resonant neutron spin-echo spectrometer RESEDA. Equivalent circuits based on these developments were installed as a spin-echo option at the triple-axis instrument MIRA \cite{MIRA}. The resonant circuits are at the heart of the spectrometer's capabilities and performance, as the achievable resolution directly depends on the maximum frequency. To ensure the design goals of a maximum frequency close to 4\,MHz and a broad bandwidth reaching as low as 35\.kHz, two separate complementary capacitor boxes were designed. The first box accommodates capacitors for low and medium frequencies as well as signal shaping electronics and transformers for impedance matching, while the second box accommodates the high frequency electronics and is therefore mounted as close as possible to the RF flipper coil to minimize additional impedance due to cable length. Only small gaps in the frequency bands were observed due to the values of the capacitors available commercially. With the new resonant circuits, a resonant $\pi$-flip was demonstrated at 3.6\,MHz even for a neutron beam with wavelength band centered at 4.5\,\AA\ where the highest current is needed.

{\btxt \subsection{Outlook}}

With the new resonating circuits in place, the performance bottleneck is now shifted to achieving higher static magnetic field $B_0$ strengths.  To resolve these limitations, {\btxt two} compact superconducting magnets with a {\btxt maximum} field strength of 300\,mT are currently being manufactured {\btxt to be installed at the primary spectrometer arm RF flippers for use in the MIEZE configuration of RESEDA}. An initial test using {\btxt a superconducting} magnet from the MLZ sample environment group with 122\,mT $B_0$ field applied at the RF flipper demonstrated the technical viability of this setup.

In addition, for MIEZE both the time and spatial resolution of the detector need to be improved. In the current configuration of RESEDA, a frequency of 3.6\,MHz leads to an intensity modulation of 2\,MHz for the neutron beam, which can only be sampled with {\btxt five or less} data points per oscillation period at the current detector speed. 
{\btxt To improve upon this, the CIPix ASIC preamplifier readout of the detector electronics \cite{Baumeister1999} would need to be improved to handle frequencies beyond 10\,MHz (corresponding to a time resolution of 100\,ns)}
Furthermore, high modulation frequencies lead to a large number of densely spaced phase rings on the detector, making a high spatial resolution crucial in order not to average over substantial phase differences in each detector pixel. {\btxt \cite{2011Brandl,2018Martin}} {\btxt After the installation of the new 300\,mT superconducting B$_0$ coils it would be necessary to reduce the pixel size of the detector by a factor of 6 from the current pixel width of 1.56 mm to effectively use the entire detector area at the highest MIEZE frequencies ($f_{MIEZE}^{max}$ = 4.9\,MHz) available at a static field of 300\,mT.}

The high frequencies now available allow the detector to be translated forward while maintaining adequate resolution. This allows for coverage over a larger solid angle with the detector, which drastically reduces the measurement times for a variety of experiments. {\btxt Additionally, technological advances in the production of GEM foils for detector application \cite{Shah2019} are smoothing the way for the production of CASCADE - type detectors with a larger active area offering even more flexibility in balancing counting time vs. spatial resolution.}

{\btxt In contrast to MIEZE, in NRSE the echo group's spatial width along the neutron flight path is only proportional to the neutron wavelength-band and thus is decoupled from the actual RF-flipper frequencies. This means that in NRSE the bottlenecks from the detector are count-rate and Q-resolution, but not time and spatial resolution.}

{\btxt Adopting the resonance spin echo techniques MIEZE and NRSE to spallation sources only requires a synchronous modulation of the RF-field amplitudes, i.e. the amplitude modulation of the sinusoidal signals in the RF generators are triggered by the pulsed source. Experiments \cite{brandl2012tests, PhysRevApplied.14.054032} and a proposal \cite{georgii2016respect} for such a LNRSE/MIEZE instrument at pulsed sources have already been reported in anticipation of ESS.}\\

\section{Acknowledgements}
We wish to thank Thomas Keller (MPI-FKF), Thomas Rapp (TUM, E10), Peter B\"oni (TUM, E21), and Robert Georgii (MLZ) for discussions and support as well as generously sharing their insights. {\btxt O.S. acknowledges} financial support through BMBF project "Longitudinale Resonante Neutronen Spin-Echo Spektroskopie mit Extremer Energie-Auflösung" (RESEDA-Plus, 05K16WO6). J.C.L. acknowledges financial support through BMBF project "Resonante Longitudinale MISANS Spin-Echo Spektroskopie an RESEDA" (MIASANS, 05K19WO5).\\

\bibliography{bibliography}

\begin{thebibliography}{34}%
\makeatletter
\providecommand \@ifxundefined [1]{%
 \@ifx{#1\undefined}
}%
\providecommand \@ifnum [1]{%
 \ifnum #1\expandafter \@firstoftwo
 \else \expandafter \@secondoftwo
 \fi
}%
\providecommand \@ifx [1]{%
 \ifx #1\expandafter \@firstoftwo
 \else \expandafter \@secondoftwo
 \fi
}%
\providecommand \natexlab [1]{#1}%
\providecommand \enquote  [1]{``#1''}%
\providecommand \bibnamefont  [1]{#1}%
\providecommand \bibfnamefont [1]{#1}%
\providecommand \citenamefont [1]{#1}%
\providecommand \href@noop [0]{\@secondoftwo}%
\providecommand \href [0]{\begingroup \@sanitize@url \@href}%
\providecommand \@href[1]{\@@startlink{#1}\@@href}%
\providecommand \@@href[1]{\endgroup#1\@@endlink}%
\providecommand \@sanitize@url [0]{\catcode `\\12\catcode `\$12\catcode
  `\&12\catcode `\#12\catcode `\^12\catcode `\_12\catcode `\%12\relax}%
\providecommand \@@startlink[1]{}%
\providecommand \@@endlink[0]{}%
\providecommand \url  [0]{\begingroup\@sanitize@url \@url }%
\providecommand \@url [1]{\endgroup\@href {#1}{\urlprefix }}%
\providecommand \urlprefix  [0]{URL }%
\providecommand \Eprint [0]{\href }%
\providecommand \doibase [0]{https://doi.org/}%
\providecommand \selectlanguage [0]{\@gobble}%
\providecommand \bibinfo  [0]{\@secondoftwo}%
\providecommand \bibfield  [0]{\@secondoftwo}%
\providecommand \translation [1]{[#1]}%
\providecommand \BibitemOpen [0]{}%
\providecommand \bibitemStop [0]{}%
\providecommand \bibitemNoStop [0]{.\EOS\space}%
\providecommand \EOS [0]{\spacefactor3000\relax}%
\providecommand \BibitemShut  [1]{\csname bibitem#1\endcsname}%
\let\auto@bib@innerbib\@empty
\bibitem [{\citenamefont {Mezei}(1972)}]{1972Mezei}%
  \BibitemOpen
  \bibfield  {author} {\bibinfo {author} {\bibfnamefont {F.}~\bibnamefont
  {Mezei}},\ }\bibfield  {title} {\bibinfo {title} {{Neutron spin echo: A new
  concept in polarized thermal neutron techniques}},\ }\href@noop {} {\bibfield
   {journal} {\bibinfo  {journal} {Zeitschrift f{\"u}r Physik A - Hadrons \&
  Nuclei}\ }\textbf {\bibinfo {volume} {255}},\ \bibinfo {pages} {146}
  (\bibinfo {year} {1972})}\BibitemShut {NoStop}%
\bibitem [{\citenamefont {Mezei}(1980)}]{1980Mezei}%
  \BibitemOpen
  \bibfield  {author} {\bibinfo {author} {\bibfnamefont {F.}~\bibnamefont
  {Mezei}},\ }\bibfield  {title} {\bibinfo {title} {{The principles of neutron
  spin echo}},\ }in\ \href@noop {} {\emph {\bibinfo {booktitle} {Neutron Spin
  Echo. Lecture Notes in Physics}}},\ Vol.\ \bibinfo {volume} {128}\ (\bibinfo
  {publisher} {Springer, Berlin, Heidelberg},\ \bibinfo {year}
  {1980})\BibitemShut {NoStop}%
\bibitem [{\citenamefont {Takeda}\ \emph {et~al.}(1995)\citenamefont {Takeda},
  \citenamefont {Komura}, \citenamefont {Seto}, \citenamefont {Nagai},
  \citenamefont {Kobayashi}, \citenamefont {Yokoi}, \citenamefont {Zeyen},
  \citenamefont {Ebisawa}, \citenamefont {Tasaki}, \citenamefont {Ito},
  \citenamefont {Takahashi},\ and\ \citenamefont
  {Yoshizawa}}]{B_int_reference}%
  \BibitemOpen
  \bibfield  {author} {\bibinfo {author} {\bibfnamefont {T.}~\bibnamefont
  {Takeda}}, \bibinfo {author} {\bibfnamefont {S.}~\bibnamefont {Komura}},
  \bibinfo {author} {\bibfnamefont {H.}~\bibnamefont {Seto}}, \bibinfo {author}
  {\bibfnamefont {M.}~\bibnamefont {Nagai}}, \bibinfo {author} {\bibfnamefont
  {H.}~\bibnamefont {Kobayashi}}, \bibinfo {author} {\bibfnamefont
  {E.}~\bibnamefont {Yokoi}}, \bibinfo {author} {\bibfnamefont {C.~M.}\
  \bibnamefont {Zeyen}}, \bibinfo {author} {\bibfnamefont {T.}~\bibnamefont
  {Ebisawa}}, \bibinfo {author} {\bibfnamefont {S.}~\bibnamefont {Tasaki}},
  \bibinfo {author} {\bibfnamefont {Y.}~\bibnamefont {Ito}}, \bibinfo {author}
  {\bibfnamefont {S.}~\bibnamefont {Takahashi}},\ and\ \bibinfo {author}
  {\bibfnamefont {H.}~\bibnamefont {Yoshizawa}},\ }\bibfield  {title} {\bibinfo
  {title} {A neutron spin echo spectrometer with two optimal field shape coils
  for neutron spin precession},\ }\href
  {https://doi.org/https://doi.org/10.1016/0168-9002(95)00336-3} {\bibfield
  {journal} {\bibinfo  {journal} {Nuclear Instruments and Methods in Physics
  Research Section A: Accelerators, Spectrometers, Detectors and Associated
  Equipment}\ }\textbf {\bibinfo {volume} {364}},\ \bibinfo {pages} {186 }
  (\bibinfo {year} {1995})}\BibitemShut {NoStop}%
\bibitem [{\citenamefont {Farago}(1999)}]{FARAGO1999270}%
  \BibitemOpen
  \bibfield  {author} {\bibinfo {author} {\bibfnamefont {B.}~\bibnamefont
  {Farago}},\ }\bibfield  {title} {\bibinfo {title} {Recent neutron spin-echo
  developments at the ill (in11 and in15)},\ }\href
  {https://doi.org/https://doi.org/10.1016/S0921-4526(99)00109-X} {\bibfield
  {journal} {\bibinfo  {journal} {Physica B: Condensed Matter}\ }\textbf
  {\bibinfo {volume} {267-268}},\ \bibinfo {pages} {270 } (\bibinfo {year}
  {1999})}\BibitemShut {NoStop}%
\bibitem [{\citenamefont {Farago}\ \emph {et~al.}(2015)\citenamefont {Farago},
  \citenamefont {Falus}, \citenamefont {Hoffmann}, \citenamefont {Gradzielski},
  \citenamefont {Thomas},\ and\ \citenamefont {Gomez}}]{IN15}%
  \BibitemOpen
  \bibfield  {author} {\bibinfo {author} {\bibfnamefont {B.}~\bibnamefont
  {Farago}}, \bibinfo {author} {\bibfnamefont {P.}~\bibnamefont {Falus}},
  \bibinfo {author} {\bibfnamefont {I.}~\bibnamefont {Hoffmann}}, \bibinfo
  {author} {\bibfnamefont {M.}~\bibnamefont {Gradzielski}}, \bibinfo {author}
  {\bibfnamefont {F.}~\bibnamefont {Thomas}},\ and\ \bibinfo {author}
  {\bibfnamefont {C.}~\bibnamefont {Gomez}},\ }\bibfield  {title} {\bibinfo
  {title} {The in15 upgrade},\ }\href
  {https://doi.org/https://doi.org/10.1080/10448632.2015.1057052} {\bibfield
  {journal} {\bibinfo  {journal} {Neutron News}\ }\textbf {\bibinfo {volume}
  {26}},\ \bibinfo {pages} {15} (\bibinfo {year} {2015})}\BibitemShut {NoStop}%
\bibitem [{\citenamefont {Pasini}\ \emph {et~al.}(2019)\citenamefont {Pasini},
  \citenamefont {Holderer}, \citenamefont {Kozielewski}, \citenamefont
  {Richter},\ and\ \citenamefont {Monkenbusch}}]{Pasini2019}%
  \BibitemOpen
  \bibfield  {author} {\bibinfo {author} {\bibfnamefont {S.}~\bibnamefont
  {Pasini}}, \bibinfo {author} {\bibfnamefont {O.}~\bibnamefont {Holderer}},
  \bibinfo {author} {\bibfnamefont {T.}~\bibnamefont {Kozielewski}}, \bibinfo
  {author} {\bibfnamefont {D.}~\bibnamefont {Richter}},\ and\ \bibinfo {author}
  {\bibfnamefont {M.}~\bibnamefont {Monkenbusch}},\ }\bibfield  {title}
  {\bibinfo {title} {J-{NSE}-phoenix, a neutron spin-echo spectrometer with
  optimized superconducting precession coils at the {MLZ} in garching},\ }\href
  {https://doi.org/10.1063/1.5084303} {\bibfield  {journal} {\bibinfo
  {journal} {Review of Scientific Instruments}\ }\textbf {\bibinfo {volume}
  {90}},\ \bibinfo {pages} {043107} (\bibinfo {year} {2019})}\BibitemShut
  {NoStop}%
\bibitem [{\citenamefont {Pasini}\ and\ \citenamefont
  {Monkenbusch}(2015)}]{Pasini2015}%
  \BibitemOpen
  \bibfield  {author} {\bibinfo {author} {\bibfnamefont {S.}~\bibnamefont
  {Pasini}}\ and\ \bibinfo {author} {\bibfnamefont {M.}~\bibnamefont
  {Monkenbusch}},\ }\bibfield  {title} {\bibinfo {title} {Optimized
  superconducting coils for a high-resolution neutron spin-echo spectrometer at
  the european spallation source, {ESS}},\ }\href
  {https://doi.org/10.1088/0957-0233/26/3/035501} {\bibfield  {journal}
  {\bibinfo  {journal} {Measurement Science and Technology}\ }\textbf {\bibinfo
  {volume} {26}},\ \bibinfo {pages} {035501} (\bibinfo {year}
  {2015})}\BibitemShut {NoStop}%
\bibitem [{\citenamefont {G{\"a}hler}\ \emph {et~al.}(1992)\citenamefont
  {G{\"a}hler}, \citenamefont {Golub},\ and\ \citenamefont
  {Keller}}]{1992Gaehler}%
  \BibitemOpen
  \bibfield  {author} {\bibinfo {author} {\bibfnamefont {R.}~\bibnamefont
  {G{\"a}hler}}, \bibinfo {author} {\bibfnamefont {R.}~\bibnamefont {Golub}},\
  and\ \bibinfo {author} {\bibfnamefont {T.}~\bibnamefont {Keller}},\
  }\bibfield  {title} {\bibinfo {title} {{Neutron resonance spin echo -- a new
  tool for high resolution spectroscopy}},\ }\href@noop {} {\bibfield
  {journal} {\bibinfo  {journal} {Physica B: Condensed Matter}\ }\textbf
  {\bibinfo {volume} {180}},\ \bibinfo {pages} {899} (\bibinfo {year}
  {1992})}\BibitemShut {NoStop}%
\bibitem [{\citenamefont {Cook}(2014)}]{2014Cook}%
  \BibitemOpen
  \bibfield  {author} {\bibinfo {author} {\bibfnamefont {J.~C.}\ \bibnamefont
  {Cook}},\ }\bibfield  {title} {\bibinfo {title} {{Concepts \& Engineering
  Aspects of a Neutron Resonance Spin-Echo Spectrometer for the National
  Institute of Standards and Technology Center for Neutron Research}},\
  }\href@noop {} {\bibfield  {journal} {\bibinfo  {journal} {Journal of
  Research}\ }\textbf {\bibinfo {volume} {119}},\ \bibinfo {pages} {55}
  (\bibinfo {year} {2014})}\BibitemShut {NoStop}%
\bibitem [{\citenamefont {Krautloher}\ \emph {et~al.}(2016)\citenamefont
  {Krautloher}, \citenamefont {Kindervater}, \citenamefont {Keller},\ and\
  \citenamefont {H{\"a}ussler}}]{2016Krautloher}%
  \BibitemOpen
  \bibfield  {author} {\bibinfo {author} {\bibfnamefont {M.}~\bibnamefont
  {Krautloher}}, \bibinfo {author} {\bibfnamefont {J.}~\bibnamefont
  {Kindervater}}, \bibinfo {author} {\bibfnamefont {T.}~\bibnamefont
  {Keller}},\ and\ \bibinfo {author} {\bibfnamefont {W.}~\bibnamefont
  {H{\"a}ussler}},\ }\bibfield  {title} {\bibinfo {title} {{Neutron resonance
  spin echo with longitudinal DC fields}},\ }\href@noop {} {\bibfield
  {journal} {\bibinfo  {journal} {Review of Scientific Instruments}\ }\textbf
  {\bibinfo {volume} {87}},\ \bibinfo {pages} {125110} (\bibinfo {year}
  {2016})}\BibitemShut {NoStop}%
\bibitem [{\citenamefont {Jochum}\ \emph {et~al.}(2019)\citenamefont {Jochum},
  \citenamefont {Wendl}, \citenamefont {Keller},\ and\ \citenamefont
  {Franz}}]{jochum2019neutron}%
  \BibitemOpen
  \bibfield  {author} {\bibinfo {author} {\bibfnamefont {J.}~\bibnamefont
  {Jochum}}, \bibinfo {author} {\bibfnamefont {A.}~\bibnamefont {Wendl}},
  \bibinfo {author} {\bibfnamefont {T.}~\bibnamefont {Keller}},\ and\ \bibinfo
  {author} {\bibfnamefont {C.}~\bibnamefont {Franz}},\ }\bibfield  {title}
  {\bibinfo {title} {Neutron mieze spectroscopy with focal length tuning},\
  }\href@noop {} {\bibfield  {journal} {\bibinfo  {journal} {Measurement
  Science and Technology}\ }\textbf {\bibinfo {volume} {31}},\ \bibinfo {pages}
  {035902} (\bibinfo {year} {2019})}\BibitemShut {NoStop}%
\bibitem [{\citenamefont {K{\"o}ppe}\ \emph {et~al.}(1999)\citenamefont
  {K{\"o}ppe}, \citenamefont {Bleuel}, \citenamefont {G{\"a}hler},
  \citenamefont {Golub}, \citenamefont {Hank}, \citenamefont {Keller},
  \citenamefont {Longeville}, \citenamefont {Rauch},\ and\ \citenamefont
  {Wuttke}}]{1999Koppe}%
  \BibitemOpen
  \bibfield  {author} {\bibinfo {author} {\bibfnamefont {M.}~\bibnamefont
  {K{\"o}ppe}}, \bibinfo {author} {\bibfnamefont {M.}~\bibnamefont {Bleuel}},
  \bibinfo {author} {\bibfnamefont {R.}~\bibnamefont {G{\"a}hler}}, \bibinfo
  {author} {\bibfnamefont {R.}~\bibnamefont {Golub}}, \bibinfo {author}
  {\bibfnamefont {P.}~\bibnamefont {Hank}}, \bibinfo {author} {\bibfnamefont
  {T.}~\bibnamefont {Keller}}, \bibinfo {author} {\bibfnamefont
  {S.}~\bibnamefont {Longeville}}, \bibinfo {author} {\bibfnamefont
  {U.}~\bibnamefont {Rauch}},\ and\ \bibinfo {author} {\bibfnamefont
  {J.}~\bibnamefont {Wuttke}},\ }\bibfield  {title} {\bibinfo {title}
  {{Prospects of resonance spin echo}},\ }\href@noop {} {\bibfield  {journal}
  {\bibinfo  {journal} {Physica B: Condensed Matter}\ }\textbf {\bibinfo
  {volume} {266}},\ \bibinfo {pages} {75} (\bibinfo {year} {1999})}\BibitemShut
  {NoStop}%
\bibitem [{\citenamefont {H{\"a}ussler}\ \emph {et~al.}(2011)\citenamefont
  {H{\"a}ussler}, \citenamefont {B{\"o}ni}, \citenamefont {Klein},
  \citenamefont {Schmidt}, \citenamefont {Schmidt}, \citenamefont {Groitl},\
  and\ \citenamefont {Kindervater}}]{2011Haeussler}%
  \BibitemOpen
  \bibfield  {author} {\bibinfo {author} {\bibfnamefont {W.}~\bibnamefont
  {H{\"a}ussler}}, \bibinfo {author} {\bibfnamefont {P.}~\bibnamefont
  {B{\"o}ni}}, \bibinfo {author} {\bibfnamefont {M.}~\bibnamefont {Klein}},
  \bibinfo {author} {\bibfnamefont {C.~J.}\ \bibnamefont {Schmidt}}, \bibinfo
  {author} {\bibfnamefont {U.}~\bibnamefont {Schmidt}}, \bibinfo {author}
  {\bibfnamefont {F.}~\bibnamefont {Groitl}},\ and\ \bibinfo {author}
  {\bibfnamefont {J.}~\bibnamefont {Kindervater}},\ }\bibfield  {title}
  {\bibinfo {title} {{Detection of high frequency intensity oscillations at
  RESEDA using the CASCADE detector}},\ }\href@noop {} {\bibfield  {journal}
  {\bibinfo  {journal} {Review of Scientific Instruments}\ }\textbf {\bibinfo
  {volume} {82}},\ \bibinfo {pages} {045101} (\bibinfo {year}
  {2011})}\BibitemShut {NoStop}%
\bibitem [{\citenamefont {Franz}\ \emph {et~al.}(2019)\citenamefont {Franz},
  \citenamefont {Soltwedel}, \citenamefont {Fuchs}, \citenamefont
  {S{\"a}ubert}, \citenamefont {Haslbeck}, \citenamefont {Wendl}, \citenamefont
  {Jochum}, \citenamefont {B{\"o}ni},\ and\ \citenamefont
  {Pfleiderer}}]{RESEDA_paper_2019}%
  \BibitemOpen
  \bibfield  {author} {\bibinfo {author} {\bibfnamefont {C.}~\bibnamefont
  {Franz}}, \bibinfo {author} {\bibfnamefont {O.}~\bibnamefont {Soltwedel}},
  \bibinfo {author} {\bibfnamefont {C.}~\bibnamefont {Fuchs}}, \bibinfo
  {author} {\bibfnamefont {S.}~\bibnamefont {S{\"a}ubert}}, \bibinfo {author}
  {\bibfnamefont {F.}~\bibnamefont {Haslbeck}}, \bibinfo {author}
  {\bibfnamefont {A.}~\bibnamefont {Wendl}}, \bibinfo {author} {\bibfnamefont
  {J.}~\bibnamefont {Jochum}}, \bibinfo {author} {\bibfnamefont
  {P.}~\bibnamefont {B{\"o}ni}},\ and\ \bibinfo {author} {\bibfnamefont
  {C.}~\bibnamefont {Pfleiderer}},\ }\bibfield  {title} {\bibinfo {title} {The
  longitudinal neutron resonant spin echo spectrometer reseda},\ }\href@noop {}
  {\bibfield  {journal} {\bibinfo  {journal} {Nuclear Instruments and Methods
  in Physics Research Section A: Accelerators, Spectrometers, Detectors and
  Associated Equipment}\ }\textbf {\bibinfo {volume} {939}},\ \bibinfo {pages}
  {22} (\bibinfo {year} {2019})}\BibitemShut {NoStop}%
\bibitem [{\citenamefont {Bloch}\ and\ \citenamefont
  {Siegert}(1940)}]{1940Bloch}%
  \BibitemOpen
  \bibfield  {author} {\bibinfo {author} {\bibfnamefont {F.}~\bibnamefont
  {Bloch}}\ and\ \bibinfo {author} {\bibfnamefont {A.}~\bibnamefont
  {Siegert}},\ }\bibfield  {title} {\bibinfo {title} {{Magnetic Resonance for
  Nonrotating Fields}},\ }\href@noop {} {\bibfield  {journal} {\bibinfo
  {journal} {Physical Review}\ }\textbf {\bibinfo {volume} {57}},\ \bibinfo
  {pages} {522} (\bibinfo {year} {1940})}\BibitemShut {NoStop}%
\bibitem [{\citenamefont {Castelnovo}\ \emph {et~al.}(2008)\citenamefont
  {Castelnovo}, \citenamefont {Moessner},\ and\ \citenamefont
  {Sondhi}}]{Castelnovo2008}%
  \BibitemOpen
  \bibfield  {author} {\bibinfo {author} {\bibfnamefont {C.}~\bibnamefont
  {Castelnovo}}, \bibinfo {author} {\bibfnamefont {R.}~\bibnamefont
  {Moessner}},\ and\ \bibinfo {author} {\bibfnamefont {S.~L.}\ \bibnamefont
  {Sondhi}},\ }\bibfield  {title} {\bibinfo {title} {Magnetic monopoles in spin
  ice},\ }\href {https://doi.org/10.1038/nature06433} {\bibfield  {journal}
  {\bibinfo  {journal} {Nature}\ }\textbf {\bibinfo {volume} {451}},\ \bibinfo
  {pages} {42} (\bibinfo {year} {2008})}\BibitemShut {NoStop}%
\bibitem [{\citenamefont {Wendl}(2018)}]{wendl2018neutron}%
  \BibitemOpen
  \bibfield  {author} {\bibinfo {author} {\bibfnamefont {A.}~\bibnamefont
  {Wendl}},\ }\emph {\bibinfo {title} {Neutron Spin Echo Spectroscopy on
  GeometricallyFrustrated Magnets}},\ \href
  {https://impulse.mlz-garching.de/record/201692} {Master's thesis} (\bibinfo
  {year} {2018})\BibitemShut {NoStop}%
\bibitem [{\citenamefont {Kindervater}\ \emph {et~al.}(2019)\citenamefont
  {Kindervater}, \citenamefont {Stasinopoulos}, \citenamefont {Bauer},
  \citenamefont {Haslbeck}, \citenamefont {Rucker}, \citenamefont {Chacon},
  \citenamefont {M\"uhlbauer}, \citenamefont {Franz}, \citenamefont {Garst},
  \citenamefont {Grundler},\ and\ \citenamefont
  {Pfleiderer}}]{Kindervater_PRX_2019}%
  \BibitemOpen
  \bibfield  {author} {\bibinfo {author} {\bibfnamefont {J.}~\bibnamefont
  {Kindervater}}, \bibinfo {author} {\bibfnamefont {I.}~\bibnamefont
  {Stasinopoulos}}, \bibinfo {author} {\bibfnamefont {A.}~\bibnamefont
  {Bauer}}, \bibinfo {author} {\bibfnamefont {F.~X.}\ \bibnamefont {Haslbeck}},
  \bibinfo {author} {\bibfnamefont {F.}~\bibnamefont {Rucker}}, \bibinfo
  {author} {\bibfnamefont {A.}~\bibnamefont {Chacon}}, \bibinfo {author}
  {\bibfnamefont {S.}~\bibnamefont {M\"uhlbauer}}, \bibinfo {author}
  {\bibfnamefont {C.}~\bibnamefont {Franz}}, \bibinfo {author} {\bibfnamefont
  {M.}~\bibnamefont {Garst}}, \bibinfo {author} {\bibfnamefont
  {D.}~\bibnamefont {Grundler}},\ and\ \bibinfo {author} {\bibfnamefont
  {C.}~\bibnamefont {Pfleiderer}},\ }\bibfield  {title} {\bibinfo {title} {Weak
  crystallization of fluctuating skyrmion textures in mnsi},\ }\href
  {https://doi.org/10.1103/PhysRevX.9.041059} {\bibfield  {journal} {\bibinfo
  {journal} {Phys. Rev. X}\ }\textbf {\bibinfo {volume} {9}},\ \bibinfo {pages}
  {041059} (\bibinfo {year} {2019})}\BibitemShut {NoStop}%
\bibitem [{\citenamefont {Martin}\ \emph {et~al.}(2014)\citenamefont {Martin},
  \citenamefont {Wagner}, \citenamefont {Dogu}, \citenamefont {Fuchs},
  \citenamefont {Kredler}, \citenamefont {Böni},\ and\ \citenamefont
  {Häußler}}]{Martin2014}%
  \BibitemOpen
  \bibfield  {author} {\bibinfo {author} {\bibfnamefont {N.}~\bibnamefont
  {Martin}}, \bibinfo {author} {\bibfnamefont {J.~N.}\ \bibnamefont {Wagner}},
  \bibinfo {author} {\bibfnamefont {M.}~\bibnamefont {Dogu}}, \bibinfo {author}
  {\bibfnamefont {C.}~\bibnamefont {Fuchs}}, \bibinfo {author} {\bibfnamefont
  {L.}~\bibnamefont {Kredler}}, \bibinfo {author} {\bibfnamefont
  {P.}~\bibnamefont {Böni}},\ and\ \bibinfo {author} {\bibfnamefont
  {W.}~\bibnamefont {Häußler}},\ }\bibfield  {title} {\bibinfo {title}
  {Neutron resonance spin flippers: Static coils manufactured by electrical
  discharge machining},\ }\href {https://doi.org/10.1063/1.4886383} {\bibfield
  {journal} {\bibinfo  {journal} {Review of Scientific Instruments}\ }\textbf
  {\bibinfo {volume} {85}},\ \bibinfo {pages} {073902} (\bibinfo {year}
  {2014})}\BibitemShut {NoStop}%
\bibitem [{\citenamefont {Geerits}\ \emph {et~al.}(2019)\citenamefont
  {Geerits}, \citenamefont {Parnell}, \citenamefont {Thijs}, \citenamefont {van
  Well}, \citenamefont {Franz}, \citenamefont {Washington}, \citenamefont
  {Raspino}, \citenamefont {Dalgliesh},\ and\ \citenamefont
  {Plomp}}]{Larmor2019}%
  \BibitemOpen
  \bibfield  {author} {\bibinfo {author} {\bibfnamefont {N.}~\bibnamefont
  {Geerits}}, \bibinfo {author} {\bibfnamefont {S.~R.}\ \bibnamefont
  {Parnell}}, \bibinfo {author} {\bibfnamefont {M.~A.}\ \bibnamefont {Thijs}},
  \bibinfo {author} {\bibfnamefont {A.~A.}\ \bibnamefont {van Well}}, \bibinfo
  {author} {\bibfnamefont {C.}~\bibnamefont {Franz}}, \bibinfo {author}
  {\bibfnamefont {A.~L.}\ \bibnamefont {Washington}}, \bibinfo {author}
  {\bibfnamefont {D.}~\bibnamefont {Raspino}}, \bibinfo {author} {\bibfnamefont
  {R.~M.}\ \bibnamefont {Dalgliesh}},\ and\ \bibinfo {author} {\bibfnamefont
  {J.}~\bibnamefont {Plomp}},\ }\bibfield  {title} {\bibinfo {title} {Time of
  flight modulation of intensity by zero effort on larmor},\ }\href
  {https://doi.org/10.1063/1.5123987} {\bibfield  {journal} {\bibinfo
  {journal} {Review of Scientific Instruments}\ }\textbf {\bibinfo {volume}
  {90}},\ \bibinfo {pages} {125101} (\bibinfo {year} {2019})}\BibitemShut
  {NoStop}%
\bibitem [{\citenamefont {Li}\ \emph {et~al.}(2014)\citenamefont {Li},
  \citenamefont {Parnell}, \citenamefont {Hamilton}, \citenamefont
  {Maranville}, \citenamefont {Wang}, \citenamefont {Semerad}, \citenamefont
  {Baxter}, \citenamefont {Cremer},\ and\ \citenamefont
  {Pynn}}]{Wollaston2014}%
  \BibitemOpen
  \bibfield  {author} {\bibinfo {author} {\bibfnamefont {F.}~\bibnamefont
  {Li}}, \bibinfo {author} {\bibfnamefont {S.~R.}\ \bibnamefont {Parnell}},
  \bibinfo {author} {\bibfnamefont {W.~A.}\ \bibnamefont {Hamilton}}, \bibinfo
  {author} {\bibfnamefont {B.~B.}\ \bibnamefont {Maranville}}, \bibinfo
  {author} {\bibfnamefont {T.}~\bibnamefont {Wang}}, \bibinfo {author}
  {\bibfnamefont {R.}~\bibnamefont {Semerad}}, \bibinfo {author} {\bibfnamefont
  {D.~V.}\ \bibnamefont {Baxter}}, \bibinfo {author} {\bibfnamefont {J.~T.}\
  \bibnamefont {Cremer}},\ and\ \bibinfo {author} {\bibfnamefont
  {R.}~\bibnamefont {Pynn}},\ }\bibfield  {title} {\bibinfo {title}
  {Superconducting magnetic wollaston prism for neutron spin encoding},\ }\href
  {https://doi.org/10.1063/1.4875984} {\bibfield  {journal} {\bibinfo
  {journal} {Review of Scientific Instruments}\ }\textbf {\bibinfo {volume}
  {85}},\ \bibinfo {pages} {053303} (\bibinfo {year} {2014})}\BibitemShut
  {NoStop}%
\bibitem [{\citenamefont {Li}\ \emph {et~al.}(2017)\citenamefont {Li},
  \citenamefont {Feng}, \citenamefont {Thaler}, \citenamefont {Parnell},
  \citenamefont {Hamilton}, \citenamefont {Crow}, \citenamefont {Yang},
  \citenamefont {Jones}, \citenamefont {Bai}, \citenamefont {Matsuda} \emph
  {et~al.}}]{Wollaston2017}%
  \BibitemOpen
  \bibfield  {author} {\bibinfo {author} {\bibfnamefont {F.}~\bibnamefont
  {Li}}, \bibinfo {author} {\bibfnamefont {H.}~\bibnamefont {Feng}}, \bibinfo
  {author} {\bibfnamefont {A.~N.}\ \bibnamefont {Thaler}}, \bibinfo {author}
  {\bibfnamefont {S.~R.}\ \bibnamefont {Parnell}}, \bibinfo {author}
  {\bibfnamefont {W.~A.}\ \bibnamefont {Hamilton}}, \bibinfo {author}
  {\bibfnamefont {L.}~\bibnamefont {Crow}}, \bibinfo {author} {\bibfnamefont
  {W.}~\bibnamefont {Yang}}, \bibinfo {author} {\bibfnamefont {A.~B.}\
  \bibnamefont {Jones}}, \bibinfo {author} {\bibfnamefont {H.}~\bibnamefont
  {Bai}}, \bibinfo {author} {\bibfnamefont {M.}~\bibnamefont {Matsuda}}, \emph
  {et~al.},\ }\bibfield  {title} {\bibinfo {title} {High resolution neutron
  larmor diffraction using superconducting magnetic wollaston prisms},\
  }\href@noop {} {\bibfield  {journal} {\bibinfo  {journal} {Scientific
  reports}\ }\textbf {\bibinfo {volume} {7}},\ \bibinfo {pages} {1} (\bibinfo
  {year} {2017})}\BibitemShut {NoStop}%
\bibitem [{\citenamefont {Dadisman}\ \emph {et~al.}(2020)\citenamefont
  {Dadisman}, \citenamefont {Wasilko}, \citenamefont {Kaiser}, \citenamefont
  {Kuhn}, \citenamefont {Buck}, \citenamefont {Schaeperkoetter}, \citenamefont
  {Crow}, \citenamefont {Riedel}, \citenamefont {Robertson}, \citenamefont
  {Jiang}, \citenamefont {Wang}, \citenamefont {Silva}, \citenamefont {Kang},
  \citenamefont {Lee}, \citenamefont {Hong},\ and\ \citenamefont
  {Li}}]{SC_RF_flipper}%
  \BibitemOpen
  \bibfield  {author} {\bibinfo {author} {\bibfnamefont {R.}~\bibnamefont
  {Dadisman}}, \bibinfo {author} {\bibfnamefont {D.}~\bibnamefont {Wasilko}},
  \bibinfo {author} {\bibfnamefont {H.}~\bibnamefont {Kaiser}}, \bibinfo
  {author} {\bibfnamefont {S.~J.}\ \bibnamefont {Kuhn}}, \bibinfo {author}
  {\bibfnamefont {Z.}~\bibnamefont {Buck}}, \bibinfo {author} {\bibfnamefont
  {J.}~\bibnamefont {Schaeperkoetter}}, \bibinfo {author} {\bibfnamefont
  {L.}~\bibnamefont {Crow}}, \bibinfo {author} {\bibfnamefont {R.}~\bibnamefont
  {Riedel}}, \bibinfo {author} {\bibfnamefont {L.}~\bibnamefont {Robertson}},
  \bibinfo {author} {\bibfnamefont {C.}~\bibnamefont {Jiang}}, \bibinfo
  {author} {\bibfnamefont {T.}~\bibnamefont {Wang}}, \bibinfo {author}
  {\bibfnamefont {N.}~\bibnamefont {Silva}}, \bibinfo {author} {\bibfnamefont
  {Y.}~\bibnamefont {Kang}}, \bibinfo {author} {\bibfnamefont {S.-W.}\
  \bibnamefont {Lee}}, \bibinfo {author} {\bibfnamefont {K.}~\bibnamefont
  {Hong}},\ and\ \bibinfo {author} {\bibfnamefont {F.}~\bibnamefont {Li}},\
  }\bibfield  {title} {\bibinfo {title} {Design and performance of a
  superconducting neutron resonance spin flipper},\ }\href
  {https://doi.org/10.1063/1.5124681} {\bibfield  {journal} {\bibinfo
  {journal} {Review of Scientific Instruments}\ }\textbf {\bibinfo {volume}
  {91}},\ \bibinfo {pages} {015117} (\bibinfo {year} {2020})}\BibitemShut
  {NoStop}%
\bibitem [{\citenamefont {Jochum}\ \emph {et~al.}(2020)\citenamefont {Jochum},
  \citenamefont {Soltwedel}, \citenamefont {Leiner},\ and\ \citenamefont
  {Franz}}]{figshare}%
  \BibitemOpen
  \bibfield  {author} {\bibinfo {author} {\bibfnamefont {J.~K.}\ \bibnamefont
  {Jochum}}, \bibinfo {author} {\bibfnamefont {O.}~\bibnamefont {Soltwedel}},
  \bibinfo {author} {\bibfnamefont {J.}~\bibnamefont {Leiner}},\ and\ \bibinfo
  {author} {\bibfnamefont {C.}~\bibnamefont {Franz}},\ }\href@noop {} {\bibinfo
  {title} {Kicad files for the circuit boards for the c-boxes of the neutron
  resonant spin echo spectrometer reseda}},\ \bibinfo {howpublished}
  {\href{https://doi.org/10.6084/m9.figshare.13177166.v1}{doi:10.6084/m9.figshare.13177166}}
  (\bibinfo {year} {2020})\BibitemShut {NoStop}%
\bibitem [{nic()}]{nicos}%
  \BibitemOpen
  \href@noop {} {\bibinfo {title} {{NICOS instrument control}}},\ \bibinfo
  {howpublished} {{https://nicos-controls.org/}}\BibitemShut {NoStop}%
\bibitem [{\citenamefont {Stunault}\ \emph {et~al.}(2006)\citenamefont
  {Stunault}, \citenamefont {Andersen}, \citenamefont {Roux}, \citenamefont
  {Bigault}, \citenamefont {Ben-Saidane},\ and\ \citenamefont
  {Rønnow}}]{Sbender_2006}%
  \BibitemOpen
  \bibfield  {author} {\bibinfo {author} {\bibfnamefont {A.}~\bibnamefont
  {Stunault}}, \bibinfo {author} {\bibfnamefont {K.}~\bibnamefont {Andersen}},
  \bibinfo {author} {\bibfnamefont {S.}~\bibnamefont {Roux}}, \bibinfo {author}
  {\bibfnamefont {T.}~\bibnamefont {Bigault}}, \bibinfo {author} {\bibfnamefont
  {K.}~\bibnamefont {Ben-Saidane}},\ and\ \bibinfo {author} {\bibfnamefont
  {H.}~\bibnamefont {Rønnow}},\ }\bibfield  {title} {\bibinfo {title} {New
  solid state polarizing bender for cold neutrons},\ }\href
  {https://doi.org/https://doi.org/10.1016/j.physb.2006.05.396} {\bibfield
  {journal} {\bibinfo  {journal} {Physica B: Condensed Matter}\ }\textbf
  {\bibinfo {volume} {385-386}},\ \bibinfo {pages} {1152 } (\bibinfo {year}
  {2006})}\BibitemShut {NoStop}%
\bibitem [{\citenamefont {Georgii}\ \emph {et~al.}(2018)\citenamefont
  {Georgii}, \citenamefont {Weber}, \citenamefont {Brandl}, \citenamefont
  {Skoulatos}, \citenamefont {Janoschek}, \citenamefont {M{\"u}hlbauer},
  \citenamefont {Pfleiderer},\ and\ \citenamefont {B{\"o}ni}}]{MIRA}%
  \BibitemOpen
  \bibfield  {author} {\bibinfo {author} {\bibfnamefont {R.}~\bibnamefont
  {Georgii}}, \bibinfo {author} {\bibfnamefont {T.}~\bibnamefont {Weber}},
  \bibinfo {author} {\bibfnamefont {G.}~\bibnamefont {Brandl}}, \bibinfo
  {author} {\bibfnamefont {M.}~\bibnamefont {Skoulatos}}, \bibinfo {author}
  {\bibfnamefont {M.}~\bibnamefont {Janoschek}}, \bibinfo {author}
  {\bibfnamefont {S.}~\bibnamefont {M{\"u}hlbauer}}, \bibinfo {author}
  {\bibfnamefont {C.}~\bibnamefont {Pfleiderer}},\ and\ \bibinfo {author}
  {\bibfnamefont {P.}~\bibnamefont {B{\"o}ni}},\ }\bibfield  {title} {\bibinfo
  {title} {{The multi-purpose three-axis spectrometer (TAS) MIRA at FRM II}},\
  }\href@noop {} {\bibfield  {journal} {\bibinfo  {journal} {Nuclear
  Instruments and Methods in Physics Research Section A: Accelerators,
  Spectrometers, Detectors and Associated Equipment}\ }\textbf {\bibinfo
  {volume} {881}} (\bibinfo {year} {2018})}\BibitemShut {NoStop}%
\bibitem [{\citenamefont {Baumeister}(1999)}]{Baumeister1999}%
  \BibitemOpen
  \bibfield  {author} {\bibinfo {author} {\bibfnamefont {D.}~\bibnamefont
  {Baumeister}},\ }\emph {\bibinfo {title} {Entwicklung und Charakterisierung
  eines ASICs zur Kathodenauslese von MWPCs f{\"u}r das H1-Experiment bei
  HERA}},\ \href@noop {} {Ph.D. thesis},\ \bibinfo  {school} {Diploma Thesis,
  Universit{\"a}t Heidelberg} (\bibinfo {year} {1999})\BibitemShut {NoStop}%
\bibitem [{\citenamefont {Brandl}\ \emph {et~al.}(2011)\citenamefont {Brandl},
  \citenamefont {Georgii}, \citenamefont {H{\"a}ussler}, \citenamefont
  {M{\"u}hlbauer},\ and\ \citenamefont {B{\"o}ni}}]{2011Brandl}%
  \BibitemOpen
  \bibfield  {author} {\bibinfo {author} {\bibfnamefont {G.}~\bibnamefont
  {Brandl}}, \bibinfo {author} {\bibfnamefont {R.}~\bibnamefont {Georgii}},
  \bibinfo {author} {\bibfnamefont {W.}~\bibnamefont {H{\"a}ussler}}, \bibinfo
  {author} {\bibfnamefont {S.}~\bibnamefont {M{\"u}hlbauer}},\ and\ \bibinfo
  {author} {\bibfnamefont {P.}~\bibnamefont {B{\"o}ni}},\ }\bibfield  {title}
  {\bibinfo {title} {{Large scales-long times: Adding high energy resolution to
  SANS}},\ }\href@noop {} {\bibfield  {journal} {\bibinfo  {journal} {Nuclear
  Instruments and Methods in Physics Research Section A: Accelerators,
  Spectrometers, Detectors and Associated Equipment}\ }\textbf {\bibinfo
  {volume} {654}},\ \bibinfo {pages} {394} (\bibinfo {year}
  {2011})}\BibitemShut {NoStop}%
\bibitem [{\citenamefont {Martin}(2018)}]{2018Martin}%
  \BibitemOpen
  \bibfield  {author} {\bibinfo {author} {\bibfnamefont {N.}~\bibnamefont
  {Martin}},\ }\bibfield  {title} {\bibinfo {title} {{On the resolution of a
  MIEZE spectrometer}},\ }\href@noop {} {\bibfield  {journal} {\bibinfo
  {journal} {Nuclear Instruments and Methods in Physics Research Section A:
  Accelerators, Spectrometers, Detectors and Associated Equipment}\ }\textbf
  {\bibinfo {volume} {882}},\ \bibinfo {pages} {11} (\bibinfo {year}
  {2018})}\BibitemShut {NoStop}%
\bibitem [{\citenamefont {Shah}\ \emph {et~al.}(2019)\citenamefont {Shah},
  \citenamefont {Sharma}, \citenamefont {Kumar}, \citenamefont {Merlin},\ and\
  \citenamefont {Naimuddin}}]{Shah2019}%
  \BibitemOpen
  \bibfield  {author} {\bibinfo {author} {\bibfnamefont {A.}~\bibnamefont
  {Shah}}, \bibinfo {author} {\bibfnamefont {A.}~\bibnamefont {Sharma}},
  \bibinfo {author} {\bibfnamefont {A.}~\bibnamefont {Kumar}}, \bibinfo
  {author} {\bibfnamefont {J.}~\bibnamefont {Merlin}},\ and\ \bibinfo {author}
  {\bibfnamefont {M.}~\bibnamefont {Naimuddin}},\ }\bibfield  {title} {\bibinfo
  {title} {Impact of single-mask hole asymmetry on the properties of gem
  detectors},\ }\href
  {https://doi.org/https://doi.org/10.1016/j.nima.2018.11.017} {\bibfield
  {journal} {\bibinfo  {journal} {Nuclear Instruments and Methods in Physics
  Research Section A: Accelerators, Spectrometers, Detectors and Associated
  Equipment}\ }\textbf {\bibinfo {volume} {936}},\ \bibinfo {pages} {459 }
  (\bibinfo {year} {2019})},\ \bibinfo {note} {frontier Detectors for Frontier
  Physics: 14th Pisa Meeting on Advanced Detectors}\BibitemShut {NoStop}%
\bibitem [{\citenamefont {Brandl}\ \emph {et~al.}(2012)\citenamefont {Brandl},
  \citenamefont {Lal}, \citenamefont {Carpenter}, \citenamefont {Crow},
  \citenamefont {Robertson}, \citenamefont {Georgii}, \citenamefont
  {B{\"o}ni},\ and\ \citenamefont {Bleuel}}]{brandl2012tests}%
  \BibitemOpen
  \bibfield  {author} {\bibinfo {author} {\bibfnamefont {G.}~\bibnamefont
  {Brandl}}, \bibinfo {author} {\bibfnamefont {J.}~\bibnamefont {Lal}},
  \bibinfo {author} {\bibfnamefont {J.}~\bibnamefont {Carpenter}}, \bibinfo
  {author} {\bibfnamefont {L.}~\bibnamefont {Crow}}, \bibinfo {author}
  {\bibfnamefont {L.}~\bibnamefont {Robertson}}, \bibinfo {author}
  {\bibfnamefont {R.}~\bibnamefont {Georgii}}, \bibinfo {author} {\bibfnamefont
  {P.}~\bibnamefont {B{\"o}ni}},\ and\ \bibinfo {author} {\bibfnamefont
  {M.}~\bibnamefont {Bleuel}},\ }\bibfield  {title} {\bibinfo {title} {Tests of
  modulated intensity small angle scattering in time of flight mode},\
  }\href@noop {} {\bibfield  {journal} {\bibinfo  {journal} {Nuclear
  Instruments and Methods in Physics Research Section A: Accelerators,
  Spectrometers, Detectors and Associated Equipment}\ }\textbf {\bibinfo
  {volume} {667}},\ \bibinfo {pages} {1} (\bibinfo {year} {2012})}\BibitemShut
  {NoStop}%
\bibitem [{\citenamefont {Oda}\ \emph {et~al.}(2020)\citenamefont {Oda},
  \citenamefont {Hino}, \citenamefont {Endo}, \citenamefont {Seto},\ and\
  \citenamefont {Kawabata}}]{PhysRevApplied.14.054032}%
  \BibitemOpen
  \bibfield  {author} {\bibinfo {author} {\bibfnamefont {T.}~\bibnamefont
  {Oda}}, \bibinfo {author} {\bibfnamefont {M.}~\bibnamefont {Hino}}, \bibinfo
  {author} {\bibfnamefont {H.}~\bibnamefont {Endo}}, \bibinfo {author}
  {\bibfnamefont {H.}~\bibnamefont {Seto}},\ and\ \bibinfo {author}
  {\bibfnamefont {Y.}~\bibnamefont {Kawabata}},\ }\bibfield  {title} {\bibinfo
  {title} {Tuning neutron resonance spin-echo spectrometers with pulsed
  beams},\ }\href {https://doi.org/10.1103/PhysRevApplied.14.054032} {\bibfield
   {journal} {\bibinfo  {journal} {Phys. Rev. Applied}\ }\textbf {\bibinfo
  {volume} {14}},\ \bibinfo {pages} {054032} (\bibinfo {year}
  {2020})}\BibitemShut {NoStop}%
\bibitem [{\citenamefont {Georgii}\ \emph {et~al.}(2016)\citenamefont
  {Georgii}, \citenamefont {Kindervater}, \citenamefont {Pfleiderer},\ and\
  \citenamefont {B{\"o}ni}}]{georgii2016respect}%
  \BibitemOpen
  \bibfield  {author} {\bibinfo {author} {\bibfnamefont {R.}~\bibnamefont
  {Georgii}}, \bibinfo {author} {\bibfnamefont {J.}~\bibnamefont
  {Kindervater}}, \bibinfo {author} {\bibfnamefont {C.}~\bibnamefont
  {Pfleiderer}},\ and\ \bibinfo {author} {\bibfnamefont {P.}~\bibnamefont
  {B{\"o}ni}},\ }\bibfield  {title} {\bibinfo {title} {Respect: Neutron
  resonance spin-echo spectrometer for extreme studies},\ }\href@noop {}
  {\bibfield  {journal} {\bibinfo  {journal} {Nuclear Instruments and Methods
  in Physics Research Section A: Accelerators, Spectrometers, Detectors and
  Associated Equipment}\ }\textbf {\bibinfo {volume} {837}},\ \bibinfo {pages}
  {123} (\bibinfo {year} {2016})}\BibitemShut {NoStop}%
\end{thebibliography}%

\newpage
\onecolumngrid 

\section{Appendix: Technical Drawings} \label{app}

\begin{figure*}[ht]
\includegraphics[width=\textwidth]{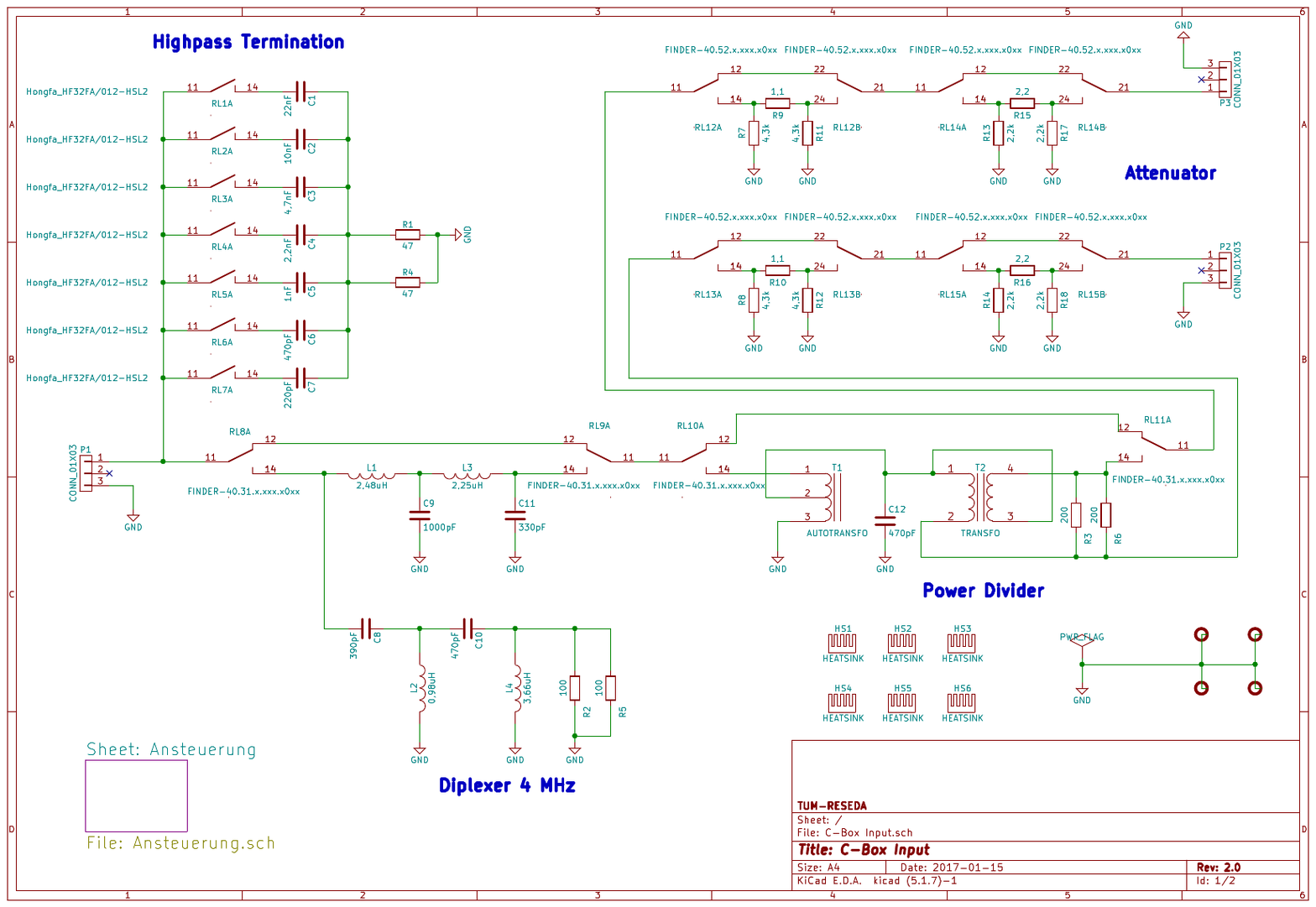}
\caption{Circuit board schematics for the input connecting to C-box1. Shown are highpass termination, diplexer and power divider.}
\label{pdf:KiCad0}
\end{figure*}

\begin{figure*}[ht]
\includegraphics[width=\textwidth, page=2]{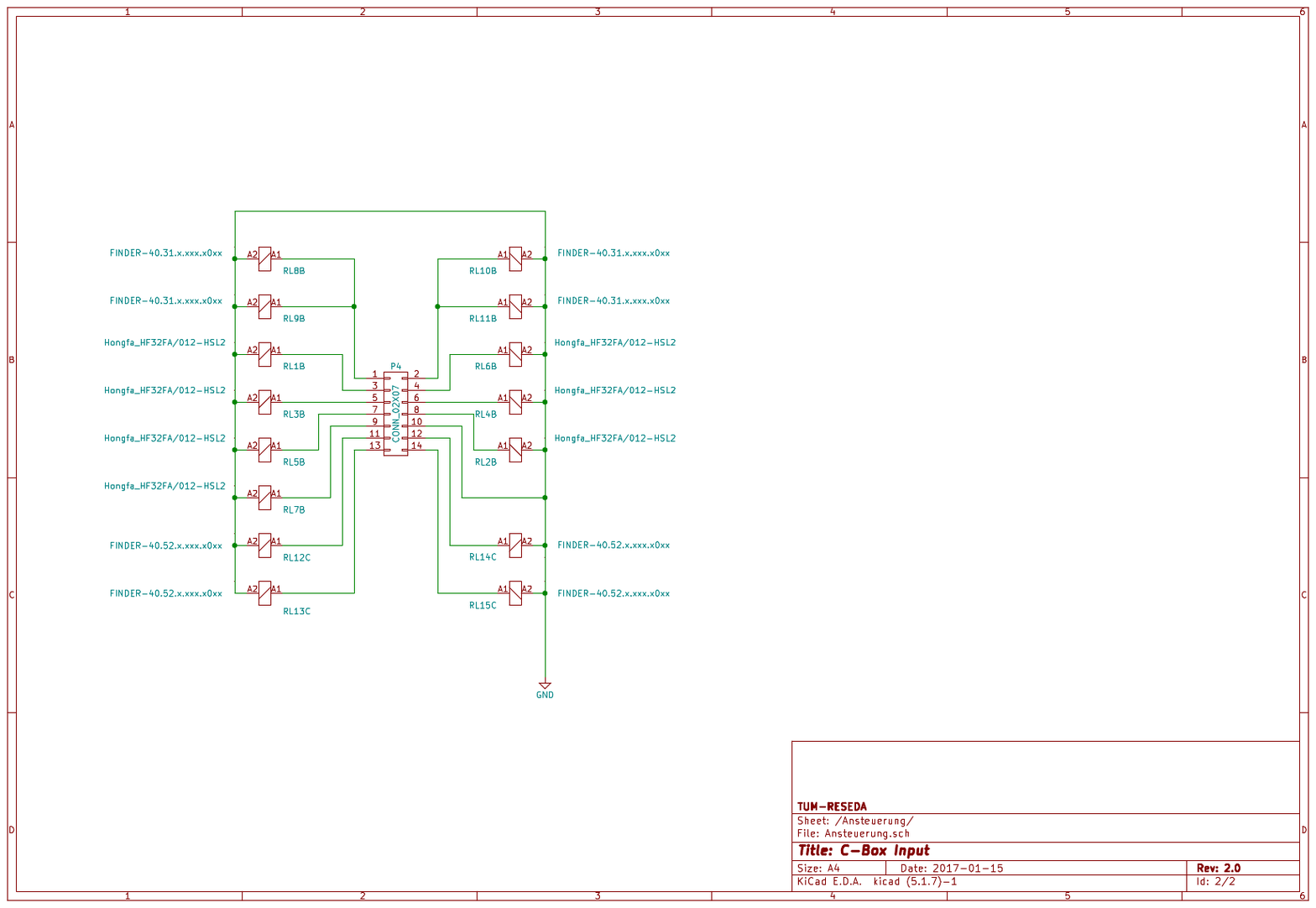}
\caption{Circuit board schematics for the input connecting to C-box1. Shown are highpass termination, diplexer and power divider.}
\label{pdf:KiCad1}
\end{figure*}

\begin{figure*}[ht]
\includegraphics[width=\textwidth]{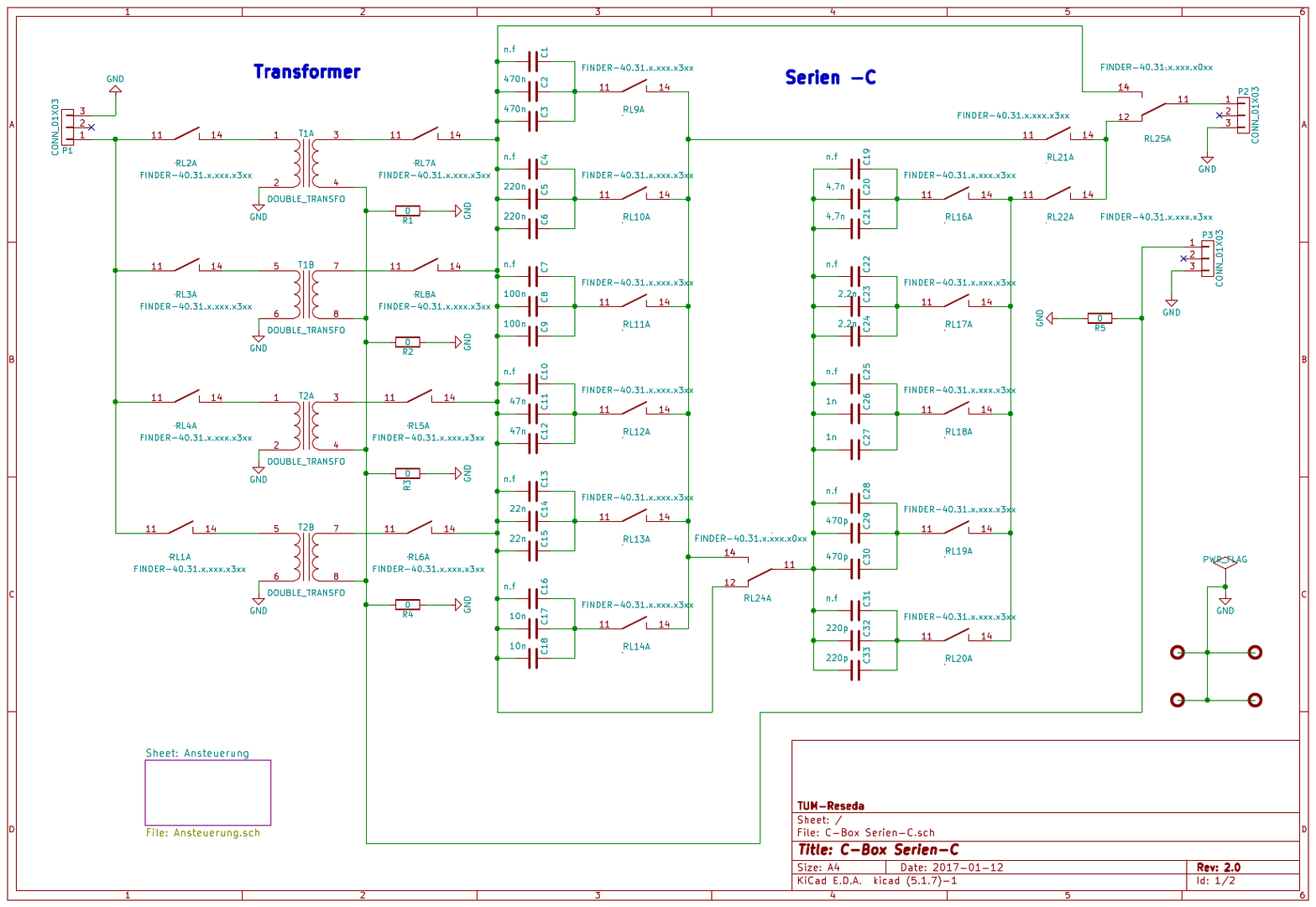}
\caption{Circuit board schematics of C-Box1, mapping out the transformers and serial capacitances used in C-Box1 in detail. }
\label{pdf:KiCad2}
\end{figure*}

\begin{figure*}[ht]
\includegraphics[width=\textwidth, page=2]{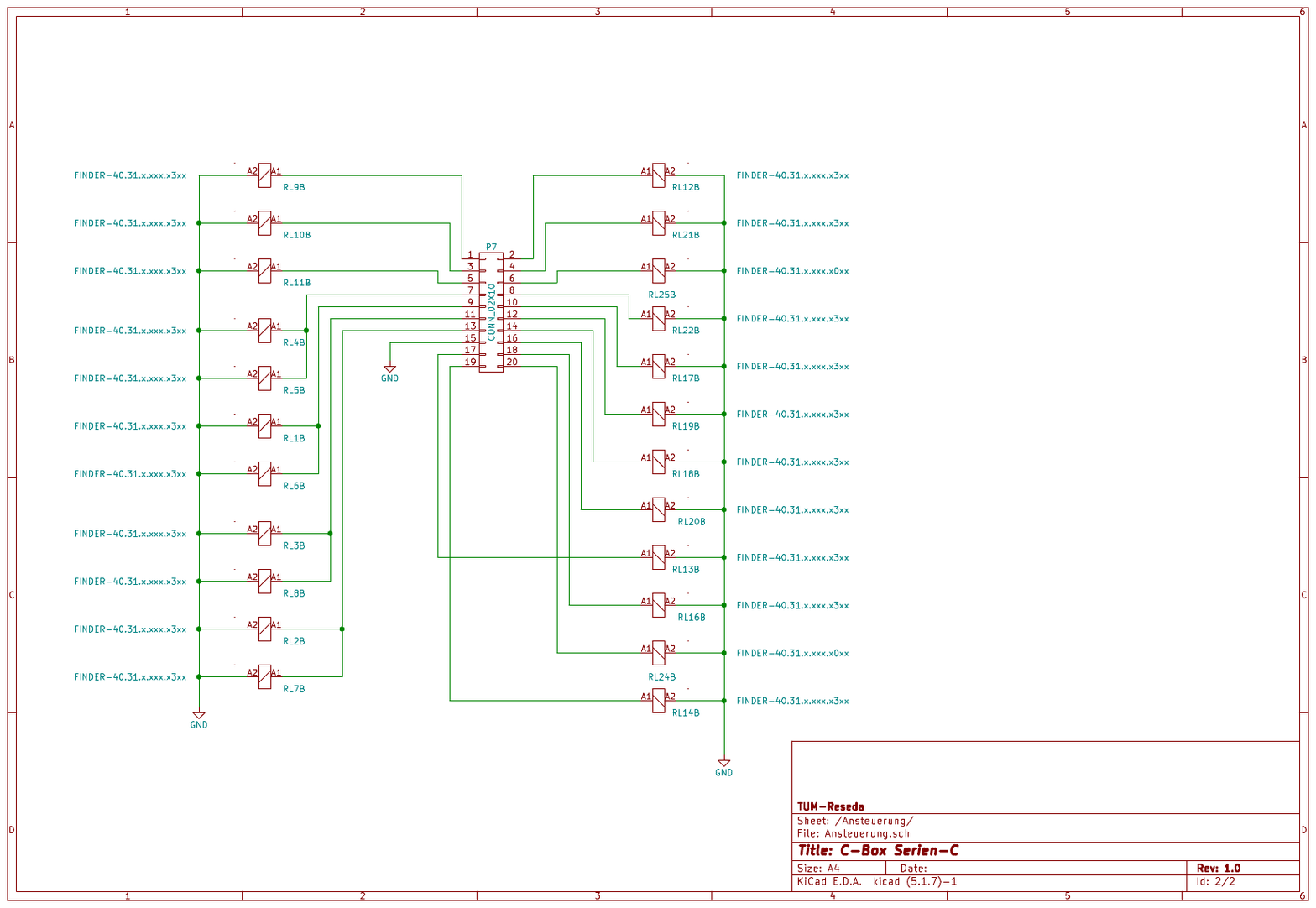}
\caption{Continuation of figure \ref{pdf:KiCad2}: Circuit board schematics of C-Box1, mapping out the transformers and serial capacitances used in C-Box1 in detail.}
\label{pdf:KiCad2p2}
\end{figure*}

\begin{figure*}[ht]
\includegraphics[width=\textwidth]{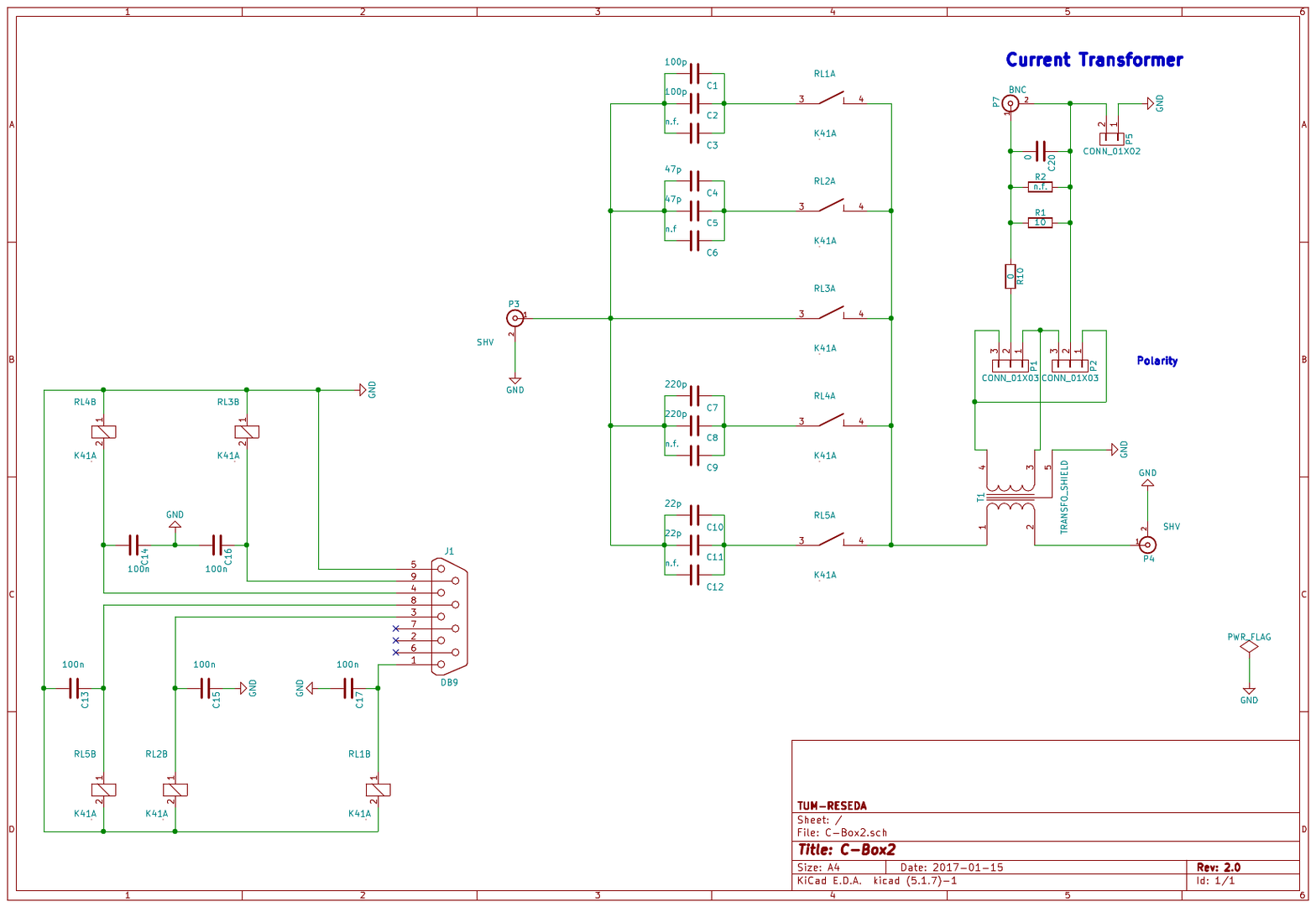}
\caption{Circuit board schematics for C-Box2. Db-9 connector to switch the relays, five high frequency relays and capacites, current transformer with filter electronics.}
\label{pdf:KiCad3}
\end{figure*}

\end{document}